\documentclass{article}

\usepackage[dblblindworkshop, final]{neurips_2026}

\usepackage[utf8]{inputenc} %
\usepackage[T1]{fontenc}    %
\usepackage{hyperref}       %
\hypersetup{
  pdftitle={MetaShare: Failure-Aware Collective Self-Improvement for Open-Ended Agents},
  pdfauthor={Vishal Bhutani, Prasang Gupta, Sumanth Chundru, Kevin Paul}
}
\usepackage{url}            %
\usepackage{booktabs}       %
\usepackage{amsfonts}       %
\usepackage{nicefrac}       %
\usepackage{microtype}      %
\usepackage{xcolor}         %
\usepackage{graphicx}
\usepackage{amsmath}
\usepackage{amssymb}
\usepackage{algorithm}
\usepackage{algpseudocode}
\usepackage{listings}
\usepackage{longtable}
\usepackage{tabularx}
\usepackage{caption}
\title{MetaShare: Failure-Aware Collective Self-Improvement for Open-Ended Agents}

\workshoptitle{Managing Agents that Manage Agents: Workshop on Responsible Use of Meta-Agents}

\author{%
  Vishal Bhutani \quad Prasang Gupta \quad Sumanth Chundru \quad Kevin Paul \\
  PricewaterhouseCoopers (PwC) \\
  \texttt{\{vishal.bhutani, prasang.gupta, sumanth.s.chundru, kevin.x.paul\}@pwc.com} \\
}

\begin{document}

\maketitle

\begin{abstract}

Humans often learn fastest not by repeating success, but by remembering which failures not to repeat. Open-ended self-improving agents can share experience across lineages, as in GEA, or edit their own improvement mechanisms, as in DGM-Hyperagents, yet failed improvement attempts are rarely retained as reusable cross-branch knowledge, and search can collapse onto redundant strategies. We introduce MetaShare, a failure-aware framework that records successful and unsuccessful agent modifications together with signed performance deltas. Before each generation, MetaShare retrieves diverse high-performing strategies alongside high-confidence failures from other lineages, letting agents reuse effective mechanisms while avoiding known dead ends. A task-blind auditor periodically injects absent or underrepresented mechanism categories, while embedding-based de-ranking limits redundancy during retrieval. Multiple meta-agents propose modifications in parallel at heterogeneous temperatures behind a deterministic selector that gates applicability, scope, compilability, and size. Experiments across paper review, coding, math grading, robotics reward design, and financial audit-risk assessment show that MetaShare's median matches or exceeds DGM-Hyperagents' final performance and reaches its own best-performing agent in 1.7--2.9$\times$ fewer evolution iterations.

\end{abstract}

\section{Introduction}

Open-ended self-improvement asks a system to keep improving the machinery it uses to solve tasks, not just its answers: a task-agent solves a problem while a meta-agent rewrites the surrounding scaffold---prompts, tools, memory, and control flow. The Darwin G\"odel Machine (DGM) and DGM-Hyperagents (DGM-H) fold task- and meta-agent into one editable, archive-based program, so the mechanism that proposes improvements can itself change \citep{zhang2025darwin,zhang2026hyperagents}; Group-Evolving Agents (GEA) widen the loop from a single branch to a population whose members inherit one another's discoveries \citep{weng2026group}. Our question is not what these systems search over, but what they carry forward from one attempt to the next. In DGM-H that is the agent, ranked by a scalar fitness; its memory can retain regressions, but only within a lineage and in whatever form an agent happened to record them. GEA broadens who inherits without changing what is inherited: members exchange agents, trajectories, and patch histories---raw material a later model must re-read before it can guess which mechanism was responsible. Credit thus lands on the agent, not the mechanism: a regressing modification leaves only implicit history, so the next meta-agent has no compact, signed signal for which mechanism to avoid and rediscovers the same dead end, while search narrows as generation and retrieval reward whatever recently scored well. What is carried forward, in short, is an agent and a number, not a reusable lesson.

\begin{figure}[t]
  \centering
  \includegraphics[width=0.72\linewidth]{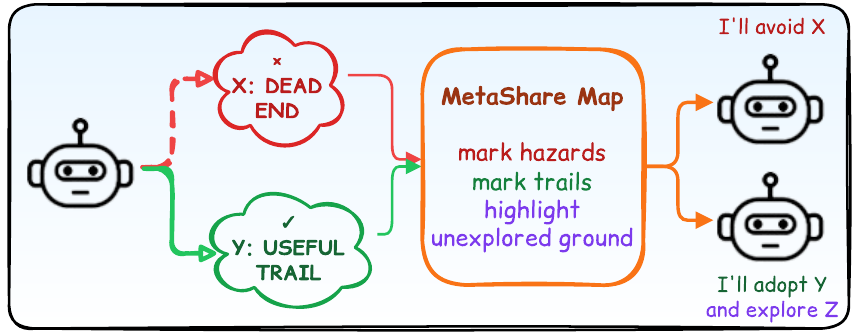}
  \caption{MetaShare converts evaluated outcomes into signed, reusable lessons. A measured scaffold change becomes a positive (helped, reuse) or negative (hurt, avoid) lesson, passed to generation $t{+}1$ to exploit gains and skip known failures.}
  \label{fig:teaser}
\end{figure}

What should it carry forward instead? Human learning suggests the useful unit is not the episode but the \emph{signed lesson}: a child who falls learns to stand and then to walk, and an adult who stumbles recovers in one motion without re-deriving why. We retain that an action helped or hurt, we retain it distilled rather than as a full trace, and we do not keep twenty copies of the same lesson. Those three properties---\emph{signed}, \emph{distilled}, and \emph{non-redundant} are precisely what a scalar fitness and a raw trace each fail to supply. Figure \ref{fig:teaser} shows an evaluated scaffold change becoming exactly such a lesson: a signed record marked \emph{helped} or \emph{hurt}, carried to generation $t{+}1$ so that gains are reused and known dead ends are skipped.

\paragraph{Our approach.} MetaShare supplies these properties by changing the unit of shared experience: each evaluated self-modification becomes a signed, mechanism-level record---a short, domain-agnostic description of the mechanism attempted (e.g.\ ``add retry logic for malformed JSON''), its signed delta, and its provenance---written only after evaluation, so it is an observed outcome rather than a self-report. Successes and failures are then retrieved through separate channels, a task-blind auditor injects absent mechanism classes, and several meta-agents propose in parallel behind a deterministic gate.

\paragraph{Contributions.}
\begin{itemize}
    \item \textbf{A signed, mechanism-level unit of shared experience.} We make the unit of cross-lineage sharing a signed record of the mechanism attempted, with an explicit avoid-channel for failures.

    \item \textbf{Task-blind search auditing.} We steer pool-level search with an auditor that targets mechanism-class coverage from a scrubbed, leak-checked digest---in contrast to task-aware reflection, it guides the meta-agent without ever seeing task labels, objectives, or absolute scores---only signed deltas.

    \item \textbf{Failure semantics, isolated.} A length-matched shuffled-failure control separates the value of a failure's \emph{content} from the value of extra context, isolating the alignment between a mechanism and its signed outcome as the source of the gain.

    \item \textbf{Cross-domain evidence.} One failure-aware mechanism improves search across five heterogeneous domains, whereas prior cross-lineage sharing was demonstrated mainly on coding.
\end{itemize}

We evaluate MetaShare against DGM-H on the same datasets, models, and evaluators across five domains---paper review, Polyglot coding, math solution grading, robotics reward design, and a new financial audit-risk task. It matches or exceeds the baseline's final quality while converging faster, and compute-matched ablations attribute the gain to the memory's structure rather than to added context or compute.

\begin{figure}[t]
  \centering
  \includegraphics[width=0.92\textwidth]{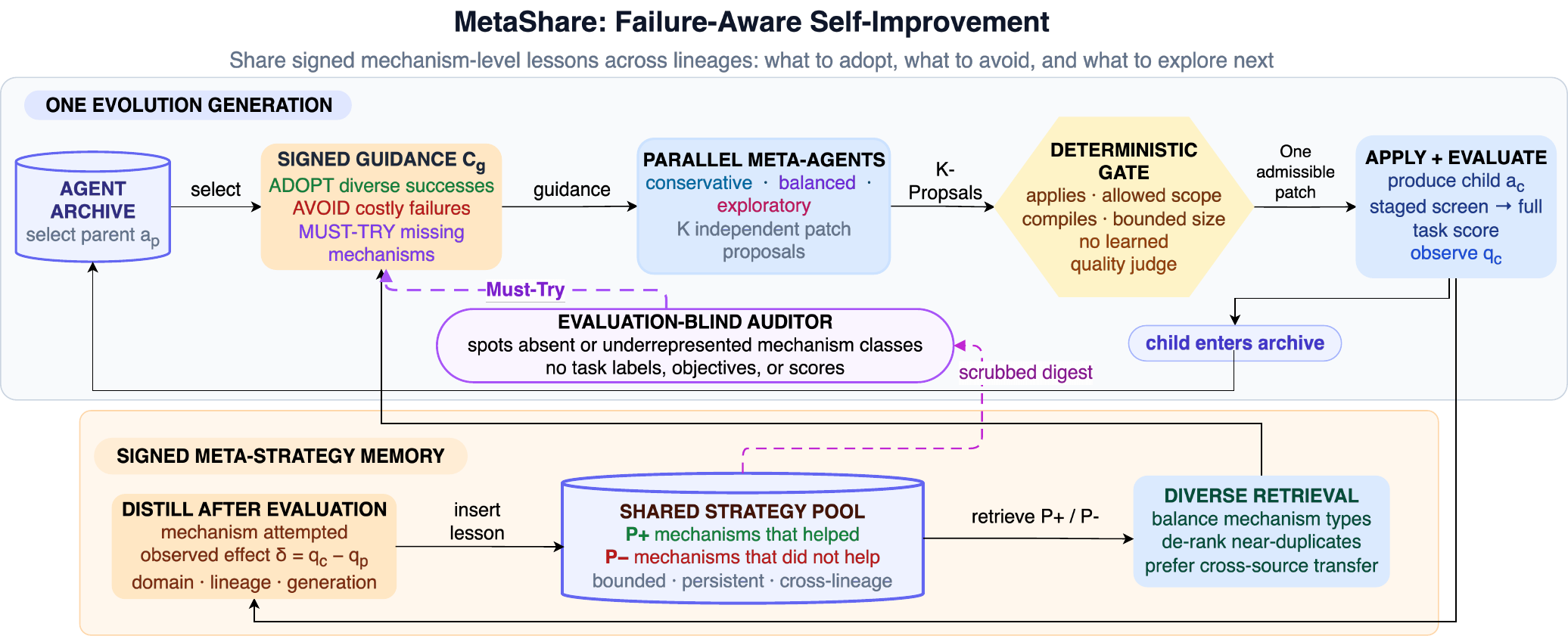}
  \caption{MetaShare architecture. \emph{Top}: one generation---select parent
  $a_p$, build the signed-guidance context $C_g$ (adopt/avoid/must-try), let
  $K$ meta-agents propose patches, admit one through a deterministic gate,
  then apply and score it to yield $a_c$. \emph{Bottom}: each evaluated
  modification becomes a signed record in a bounded cross-lineage pool;
  failure-aware diverse retrieval rebuilds $C_g$, and a task-blind auditor
  injects Must-Try classes from a scrubbed digest.}
\label{fig:architecture}
\end{figure}

\section{Related Work}

\paragraph{Open-ended self-improvement and sharing.} Self-improving systems predate foundation models \citep{good1966speculations,schmidhuber2003godel,kirsch2022eliminating}. Holding weights fixed and searching the scaffold makes this operational: STOP \citep{zelikman2024self}, ADAS \citep{hu2024automated}, DGM \citep{zhang2025darwin}, and DGM-H, which folds task- and meta-agent into one editable program \citep{zhang2026hyperagents}. GEA shares experience across a population rather than a lineage \citep{weng2026group}; ExpeL distills insights from trajectories \citep{zhao2024expel}; Reflexion and Self-Refine reuse the \emph{sign} of feedback \citep{shinn2024reflexion,madaan2023self}. MetaShare changes \emph{what} is inherited: a persistent, signed, mechanism-level record indexed by outcome across lineages, rather than an agent archive or free-form reflection.

\paragraph{Diversity-preserving search.} Novelty search rewards behavioral difference over progress toward a fixed objective \citep{lehman2011abandoning}, and quality--diversity methods such as MAP-Elites retain strong solutions across niches \citep{mouret2015illuminating}. MetaShare borrows this intuition without a QD archive: a task-blind auditor targets mechanism-category \emph{coverage} rather than ranking by score, and embedding de-ranking---related to maximal marginal relevance \citep{carbonell1998use}---reduces redundancy among retrieved modification records.

\paragraph{Parallel generation and selection.} Self-consistency shows the value of sampling and aggregating multiple reasoning paths \citep{wang2022self}; MetaShare transfers the multiple-sample idea to code modifications via parallel meta-agents at heterogeneous temperatures. Rather than a learned judge---which carries position, verbosity, and self-enhancement biases \citep{zheng2023judging}---it admits candidates by deterministic checks (applicability, scope, compilation, size), leaving quality to empirical evaluation.

\section{Method}
\label{sec:method}

We define \emph{meta-strategy memory} as the persistent storage and retrieval of evaluated improvement strategies, each summarizing the mechanism attempted, its measured effect, and its archive provenance. It is language-based because a self-modification's reusable content often lies at the mechanism level: a long diff may instantiate the same idea as ``add retry logic for malformed JSON outputs.'' Figure~\ref{fig:architecture} shows how recording (Sec.~\ref{sec:strategy-memory}), failure-aware retrieval (Sec.~\ref{sec:retrieval}), diversity control and auditing (Sec.~\ref{sec:diversity-audit}), and proposal selection (Sec.~\ref{sec:proposal-selection}) compose into one evolution generation.

\subsection{Signed Meta-Strategy Records}
\label{sec:strategy-memory}

At generation \(g\), a parent agent \(a_p=(T_p,M_p)\), with task-agent \(T_p\) and meta-agent \(M_p\), is selected from the archive \(\mathcal{A}_g\).  Given permitted history \(H_g\) and retrieved MetaShare context \(C_g\), the meta-agent proposes a patch \(p_g\), which is applied to produce a child:
\begin{equation}
    p_g=M(a_p,H_g,C_g),\qquad
    a_c=\operatorname{Apply}(a_p,p_g).
    \label{eq:child}
\end{equation}
The child transition in Eq.~\ref{eq:child} is evaluated on the domain task, and only afterward can MetaShare write to the pool: a record is thus an evaluated transition with a known outcome, not a self-report of intent. The update fires whenever a generation yields a non-empty diff and at least one valid score; it extracts the diff, computes parent and child scores, and classifies and summarizes the diff into a reusable mechanism description.

Formally, for evaluator score \(q(a;\mathcal{D})\), MetaShare converts the parent--child transition into a signed meta-strategy record:
\begin{equation}
\begin{split}
    s_i = (&\Delta_i,c_i,r_i,\delta_i,q_i^p,q_i^c,\\
           &d_i,\ell_i,g_i),\qquad
    \delta_i=q_i^c-q_i^p .
\end{split}
\label{eq:strategy-record}
\end{equation}
Here \(\Delta_i\) is the modification (task-agent, meta-agent, or both), \(c_i\) its mechanism type, \(r_i\) a concise description, \(\delta_i\) the signed performance change, \(q_i^p,q_i^c\) parent and child scores, \(d_i\) the domain, \(\ell_i\) the source-parent identifier, and \(g_i\) the generation. Type and description come from a post-hoc classification step; the type drives retrieval diversity and auditing, preventing domination by one family and letting the auditor spot underexplored classes. The fixed vocabulary is prompt engineering, tool creation, memory, evaluation logic, self-modification, and structural refactoring. Each record thus separates \emph{what} was attempted, \emph{what happened}, and \emph{where}, without exposing benchmark labels or traces.

MetaShare accumulates these records into a persistent \emph{strategy pool} \(\mathcal{P}_g=\{s_1,s_2,\dots\}\), the set of all signed records written through generation \(g\). The sign of the delta \(\delta_i\) then splits the pool into positive and non-positive evidence:

\begin{equation}
    \mathcal{P}_g^+=\{s_i\in\mathcal{P}_g:\delta_i>0\},
    \;
    \mathcal{P}_g^-=\{s_i\in\mathcal{P}_g:\delta_i\leq0\}
    \label{eq:pool-split}
\end{equation}
so \(\mathcal{P}_g^+\) holds the mechanisms that \emph{improved} performance (candidates to reuse) and \(\mathcal{P}_g^-\) holds those that did not (explicit avoid-signals). Persisting each record makes the outcome available outside its originating branch, so later meta-agents retrieve an attempt's description, sign, and provenance rather than rediscovering it. The pool is capacity-bounded at 100 records; on overflow the lowest-delta records are evicted first, so the most negative failures go before any success. Each generation writes at most one record, so only the 200-generation IMO runs can reach the bound.

\subsection{Failure-Aware Retrieval}
\label{sec:retrieval}

Before each generation, MetaShare turns the pool into a prompt context with two goals: expose the parent to mechanisms found outside its source branch, and present failures as explicit evidence, not buried log history. Since \(\ell_i\) is the source-parent identifier, filtering out records with \(\ell_i=\ell_p\) yields the \emph{eligible pool} \(\widetilde{\mathcal{P}}_g\) (with positive and non-positive parts \(\widetilde{\mathcal{P}}_g^+\) and \(\widetilde{\mathcal{P}}_g^-\)): a conservative parent-level exclusion that encourages cross-source transfer rather than full ancestry removal.

Retrieval is deliberately asymmetric: positive records are candidate mechanisms to adapt, non-positive records are constraints and warnings. From the eligible pool, MetaShare selects the two sets actually placed in the meta-agent's context---the retrieved positives \(S_g^+\) and the retrieved failures \(S_g^-\)---under a positive budget \(K_s\) and a failure budget \(K_f\):

\begin{equation}
\begin{split}
    S_g^+&=\operatorname{DiverseTopK}(\widetilde{\mathcal{P}}_g^+,K_s),\\
    S_g^-&=\operatorname{BottomK}(\widetilde{\mathcal{P}}_g^-,K_f),
\end{split}
\label{eq:retrieval-sets}
\end{equation}
Here \(\operatorname{DiverseTopK}\) returns the highest-\(\delta_i\) positives, diversified across mechanism types (Sec.~\ref{sec:diversity-audit}), and \(\operatorname{BottomK}\) returns the most negative (worst) failures; so \(S_g^+\) supplies mechanisms to \emph{adapt} and \(S_g^-\) mechanisms to \emph{avoid or revise}. By default \(K_s=5\), with up to \(K_f=5\) worst failures appended after the positives.

The formatted context preserves the sign of the evidence---positives as strategies to adopt, failures in a separate section labeled as reducing performance---so the meta-agent gets a compact mechanism description with its outcome's direction and magnitude, not a long trace to interpret.

\subsection{Diversity Control and Search Audit}
\label{sec:diversity-audit}

Failure-aware retrieval avoids repeating known bad strategies but does not prevent collapse onto a few high-scoring recipes: if recent gains come mostly from prompt edits, a delta-ranked retriever may surface several prompt variants and little on tools, memory, evaluation logic, or self-modification. MetaShare therefore adds three layers.

The first is category-aware retrieval. After ranking positives by delta, MetaShare takes at most one representative per strategy type before filling the remaining budget by score, using the type field \(c_i\) as a control variable, not a descriptive afterthought.

The second is semantic de-ranking. Each positive strategy is embedded from its description and added patch lines, with similarity measured by cosine; strategies above threshold \(\tau\) are grouped and the highest-delta member preferred. Running this at retrieval, not storage, lets the pool keep related attempts for analysis while the prompt avoids redundant paraphrases, preserving useful variation \citep{lehman2011abandoning,mouret2015illuminating,pugh2016quality} without behavioral niches or a QD score.

The third is a task-blind search audit. Every \(C\) generations it receives a scrubbed digest of strategy types, signed deltas, and added lines---task identifiers, benchmark names, and other domain-revealing patterns removed---and names absent or underrepresented mechanism classes. It does not see the task, rank agents, or solve anything.

The output is used only if it passes a leak check against the same policy---scanned for the blocked tokens and patterns above and, on detection, retried once, else rejected. A valid audit is prepended to later contexts as a \textsc{Must-Try} directive---the cached directive \(u_g\) of Algorithm~\ref{alg:metashare}---and an invalid one is discarded with the prior valid directive retained. Together the three layers separate two often-conflated goals: reusing what worked and keeping the improvement policy from narrowing too early.

\subsection{Proposal Generation, Selection, and Update}
\label{sec:proposal-selection}

After building the context, the system decides which self-modification to evaluate. MetaShare can sample several candidates for the same parent:
\begin{equation}
    p_{g,k}=M_{\theta_k}(a_p,H_g,C_g),
    \qquad k\in\{1,\ldots,K\}.
    \label{eq:parallel-proposals}
\end{equation}
Each of the \(K\) proposals runs in an independent container and repo copy; temperatures \(\theta_k\) are specified or spread uniformly over \([0.3,0.9]\). Only one is evaluated, so extra proposals raise self-modification cost but not domain-evaluation cost.

The selector is structural, not predictive. For candidate patch \(p\),
\begin{equation}
\begin{split}
    V(p) ={}& I_{\mathrm{apply}}(p)\,I_{\mathrm{scope}}(p)\,I_{\mathrm{compile}}(p)\,I_{\mathrm{size}}(p),\\
    & I_{\mathrm{size}}(p)=\mathbf{1}[\,5\leq L(p)\leq 300\,].
\end{split}
\label{eq:validity}
\end{equation}
requiring clean application, allowed scope, successful Python compilation, and a bounded line count; any failed gate disqualifies the patch, and ties are broken by a generation-seeded random choice. This keeps a learned judge out of candidate selection while evaluation measures quality; the post-hoc classifier and the auditor are language models, but neither ranks candidates.

The selected patch is applied and screened on a low-cost subset, with full evaluation run only when the staged score exceeds \(\gamma\). Values for \(\gamma\), the de-ranking threshold \(\tau\), the audit cadence \(C\), the pool capacity bound, and the embedding model are listed per domain in Appendix~\ref{sec:hyperparameters}. The evaluated child enters the archive, and the post-evaluation update turns the transition into a signed record. Algorithm~\ref{alg:metashare} summarizes the loop: retrieval builds the context, proposals generate candidates, selection picks one patch, and evaluation updates both archive and strategy memory.

\begin{algorithm}[t]
\caption{MetaShare evolution}
\label{alg:metashare}
\begin{algorithmic}[1]
\Require initial agent \(a_0\), budget \(G\), proposal count \(K\)
\State \(\mathcal{A}_0\gets\{a_0\}\), \(\mathcal{P}_0\gets\varnothing\)
\For{\(g=1,\ldots,G\)}
    \State Select parent \(a_p\) from \(\mathcal{A}_{g-1}\); load cached audit \(u_g\)
    \If{audit is due}
        \State Scrub \(\mathcal{P}_{g-1}\), run blind audit, leak-check \(u_g\)
    \EndIf
    \State Retrieve \(S_g^+,S_g^-\) from \(\mathcal{P}_{g-1}\); form context \(C_g\) with \(u_g\)
    \State Generate \(K\) patches \(\{p_{g,k}\}\); select valid \(p_g^\star\)
    \If{no valid patch}
        \State \(\mathcal{A}_g\gets\mathcal{A}_{g-1}\), \(\mathcal{P}_g\gets\mathcal{P}_{g-1}\); \textbf{continue}
    \EndIf
    \State Apply \(p_g^\star\) to get child \(a_c\); evaluate to obtain \(q_c\)
    \State \(\mathcal{A}_g\gets\mathcal{A}_{g-1}\cup\{a_c\}\)
    \State Build record \(s_g\) from \(p_g^\star,q(a_p),q_c\); insert into \(\mathcal{P}_g\)
\EndFor
\State \Return \(\mathcal{A}_G,\mathcal{P}_G\)
\end{algorithmic}
\end{algorithm}

\section{Experiments and Results}
\label{sec:experiments}

The experiments compare MetaShare against DGM-H on the same datasets, models, and evaluators, asking whether it matches or improves DGM-H's best performance \textbf{RQ1 (final quality)} and whether it reaches comparable quality in fewer generations or tokens \textbf{RQ2 (search efficiency)}. Ablations then isolate which parts of the memory and retrieval pipeline account for the gains.

\subsection{Dataset}
\label{sec:experiment-domains}

Our evaluation spans five domains: four inherited unchanged from the DGM-H suite---enabling direct comparison against the self-improving-agent baseline---and one new financial audit-risk task that we curate (described below).

\paragraph{Inherited domains.} For the four inherited domains we adopt the datasets, splits, and model configurations of DGM-H and refer the reader to \citet{zhang2026hyperagents} for their full specification, summarizing each only briefly here. \emph{Paper review} is a binary accept/reject task over research papers, testing long-context, criteria-based judgment. \emph{Polyglot coding} evaluates code generation across languages on a unique 225 coding tasks. \emph{Mathematical solution grading} (IMO) grades Olympiad-level solutions---scored by exact accuracy and normalized mean absolute error (NMAE)---in three settings: IMO~(init), IMO~(transfer), and IMO~(full). \emph{Robotics reward design} evaluates generated reward functions for quadruped locomotion by average fitness. Except for Polyglot, each uses disjoint train/validation/test splits, with the validation-selected checkpoint evaluated once on the held-out test set.

\paragraph{Financial audit-risk assessment.}
Beyond the four inherited benchmarks, we curate an audit-risk dataset from the financial disclosures of 15 real audit clients, testing whether MetaShare transfers to a document-grounded setting rather than only to the tasks it was tuned alongside.  For each client we assess the 15 Inherent Risk Factors (IRFs) defined in the audit domain, giving 225 labelled (client, IRF) pairs under a client-level split (8 train, 2 validation, 5 test), so the test fold is 75 decisions clustered within five filings.  Given chunks of a client's 10-K Financial Notes and an IRF, the task predicts one of three risk levels---High, Medium, or None---and we let the framework discover the mapping rather than supplying a hand-specified rule, so accuracy measures agreement with expert judgement rather than audit-grade assessment.  Because skewed labels can make surface-level predictions look correct, we report overall accuracy, normalized mean absolute error, and per-risk-factor accuracy; per-IRF definitions are in Appendix~\ref{sec:irf-definitions}.

\subsection{Results}
\label{sec:systems-compared}

We compare MetaShare with DGM-H on the same datasets, splits, models, and evaluators. For paper review, robotics, and the three IMO settings, DGM-H scores, intervals, and generations-to-best are those reported by \citet{zhang2026hyperagents}; we did not rerun them, so on these domains our admissibility gate (Sec.~\ref{sec:proposal-selection}) is not held fixed across arms. For Polyglot and audit risk, the baseline is our reimplementation---MetaShare with the pool disabled and a single meta-agent (\(k{=}1\)), same gate---so the difference spans the memory-and-retrieval layer and parallel proposal generation, not a change of agent or evaluator; the ablations (Sec.~\ref{sec:ablations}) attribute gains to individual components. Every condition we run is run five times, following DGM-H's protocol. GEA's released harness is coding-specific and reports no numbers on our other domains, so we compare against it only on Polyglot, under its own protocol.

\begin{figure}[t]
  \centering
  \includegraphics[width=0.8\textwidth]{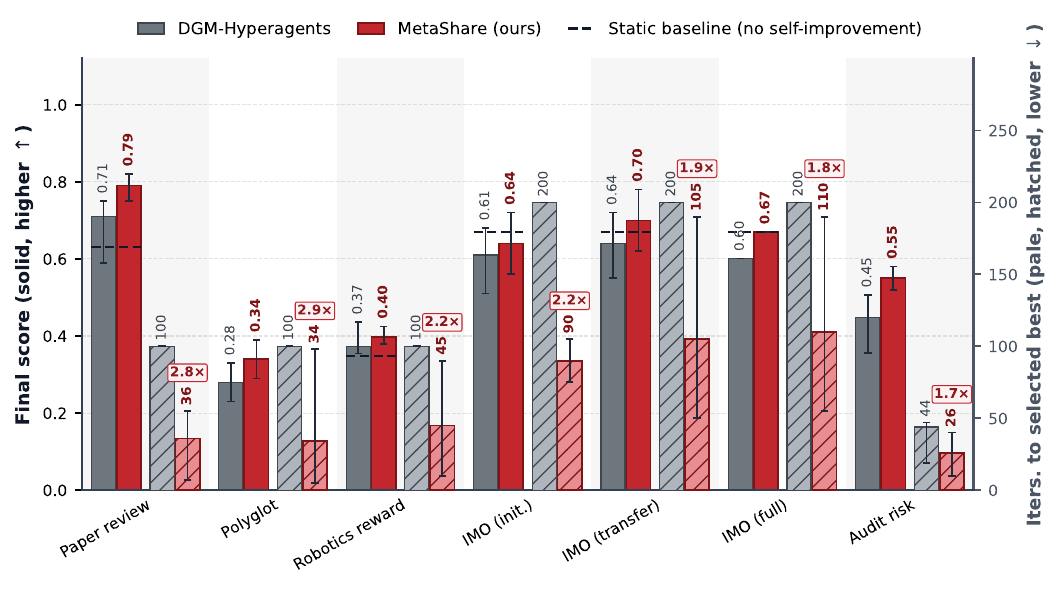}
  \caption{MetaShare vs.\ DGM-H across all evaluation domains. \emph{Left}: final
score ($\uparrow$); \emph{right}: generation of the selected best ($\downarrow$),
with per-domain speed-up. Error bars are 95\% bootstrap CIs where available.
DGM-H values for paper review, robotics, and IMO (scores, intervals, and
generations) are as reported by \citet{zhang2026hyperagents}, not rerun; the
Polyglot and audit-risk DGM-H bars are our five-run reimplementation. Dashed lines mark the static, no-self-improvement baseline
\citep{zhang2026hyperagents,luong2025towards} (none for Polyglot or audit risk).
NMAE ($\downarrow$) is $0.175$ vs.\ $0.157$ (IMO-full) and $0.357$ vs.\ $0.312$
(audit risk). For Polyglot both arms hold the task-agent on o3-mini here, with the DGM-H baseline rerun on a Sonnet-4.5 meta-agent (Sonnet-3.5 deprecated); Sec.~\ref{sec:ablation-auditor} analyzes model-routing separately.}
  \label{fig:main-comparison}
\end{figure}

We report medians over five runs with 95\% bootstrap confidence intervals, generations to the best checkpoint, and token cost. Figure~\ref{fig:main-comparison} summarizes the comparison across all five domains: MetaShare matches or exceeds DGM-H's final quality while converging in far fewer generations. Median score improves on paper review ($0.71\!\to\!0.79$), Polyglot ($0.28\!\to\!0.34$, full 225 tasks, backbone held on o3-mini), robotics reward design ($0.37\!\to\!0.40$), the three IMO-grading settings ($0.61/0.64/0.60\!\to\!0.64/0.70/0.67$), and audit risk ($0.45\!\to\!0.55$), while the generation at which the selected best appears falls by $1.7\text{--}2.9\times$ (e.g.\ $100\!\to\!36$ on paper review and $44\!\to\!26$ on audit risk). This locates each arm's best checkpoint rather than timing arrival at a fixed target, and on paper review the baseline's best coincides with its budget, so that ratio is a lower bound.

On the small and medium set of 60 tasks in Polyglot benchmark, GEA reports $0.88$ from a single run (Claude~Sonnet-4.5 as both task- and meta-agent); matching that configuration, MetaShare reaches $0.917$ [$0.900$, $0.933$] over five runs within 10 iterations. A five-run interval and a single point estimate are not like-for-like, so we read this as competitive at a lower iteration count rather than a win; per-run detail is in Appendix~\ref{sec:polyglot-runs}.

We follow DGM-H's accounting protocol \citep{zhang2026hyperagents}: only the \emph{self-modification} term differs, since the \emph{task-evaluation} term is identical per iteration. Running $k{=}3$ meta-agents makes MetaShare's self-modification about $3\times$ the single-agent baseline, but total cost stays within the same order of magnitude because evaluation dominates and is independent of $k$. Table~\ref{tab:token-cost} gives the totals: MetaShare is the more expensive method on two of the three domains we account for.

\begin{table}[t]
\small
\caption{Estimated total token count (millions, input$+$output), MetaShare (Ours) vs.\ the DGM-H baseline, following DGM-H's protocol \citep{zhang2026hyperagents}. Baseline self-modification is DGM-H's reported estimate ($\sim$33M tokens per 100 iterations, scaled to each budget); totals are medians over five runs to the selected checkpoint; Ratio is Ours$/$baseline. $^{\dagger}$Polyglot baseline re-measured on the matching model (Claude~4.5) for parity.}
\label{tab:token-cost}
\centering
\setlength{\tabcolsep}{6pt}
\begin{tabular}{lccc}
\toprule
\textbf{Domain (budget)} & \textbf{DGM-H} & \textbf{Ours} & \textbf{Ratio} \\

\midrule
Paper Review (100)          & 83M  & 52M  & $0.6\times$ \\
Polyglot (100)$^{\dagger}$  & 135M & 163M & $1.2\times$ \\
IMO-grading (200)           & 88M  & 213M & $2.4\times$ \\
\bottomrule
\end{tabular}

\end{table}

\section{Ablation Studies}
\label{sec:ablations}

The ablations test whether MetaShare's gains come from the \emph{structure} of the memory rather than from added context or compute. We ask two questions: \textbf{AQ1 (failure semantics)}---does the \emph{content} of a failure record help beyond the prompt tokens it adds?---and \textbf{AQ2 (search auditing)}---does the task-blind auditor broaden the meta-agent's search instead of letting it collapse onto redundant mechanisms? Type-diverse retrieval and semantic de-ranking are kept \emph{on} in every arm below, so each comparison holds them fixed rather than isolating them.

\subsection{Component Ladder (Paper Review)}
\label{sec:ablation-ladder}

We toggle one component at a time on paper review, holding base system, model, domain, protocol, and five paired runs fixed; Table~\ref{tab:ablation-ksweep} reports best-so-far accuracy (median, 95\% bootstrap CI) and generations-to-best. Every component helps, and the two splits agree. The signed failure channel adds \(+0.02\)/\(+0.04\) (train/val), so negative records carry usable avoid-signals, not merely extra tokens (AQ1); parallel heterogeneous proposals give the largest ladder step; and the task-blind auditor completes the system (AQ2). A \(k{=}3\) generation costs about \(3\times\) the self-modification tokens (Table~\ref{tab:token-cost}), so this is a quality-and-speed gain, not a token-efficiency claim.

The ladder adds parallel proposals partway up, so it cannot separate the memory layer from proposal count. A compute-matched reference---the pool disabled at \(k{=}3\), same gate, temperature spread, budget, and paired runs (last row of Table~\ref{tab:ablation-ksweep})---reaches \(0.72\) [\(.72\), \(.73\)] val in 55 generations. At identical proposal count the memory layer and auditor therefore account for \(+0.11\) val and 19 fewer generations, disjoint from the full system's \(0.83\) [\(.80\), \(.88\)]; signed memory alone supplies \(+0.06\) (also disjoint), the auditor \(+0.05\). The large \(k{=}1\!\to\!k{=}3\) ladder step is thus not evidence that proposal count drives the result: at matched count the memory layer accounts for the gain, and Sec.~\ref{sec:ablation-auditor} repeats the comparison on Polyglot, where proposal count alone contributes nothing.

\begin{table}[t]
\small
\caption{Component ablation on paper review: best-so-far accuracy (median over five paired runs; 95\% bootstrap CI) on train and validation---also the checkpoint-selection split---and generations-to-best on validation (Gens\,$\downarrow$). The first four rows add one component at a time; the last is the \emph{compute-matched} reference, with the pool disabled at the full system's proposal count (\(k{=}3\)).}
\label{tab:ablation-ksweep}
\centering
\setlength{\tabcolsep}{4pt}
\begin{tabular}{@{}lccc@{}}
\toprule
Configuration & Train & Val & Gens\,$\downarrow$ \\
\midrule
Success only            & 0.68 [.68,.69] & 0.69 [.69,.72] & 63 \\
\quad + failure (\(k{=}1\))   & 0.70 [.69,.70] & 0.73 [.72,.74] & 60 \\
\quad + parallel (\(k{=}3\))  & 0.77 [.76,.79] & 0.78 [.76,.83] & 48 \\
\quad + auditor (full)       & \textbf{0.81} [.79,.83] & \textbf{0.83} [.80,.88] & \textbf{36} \\
\midrule
Pool disabled, \(k{=}3\)     & 0.70[.70,.71] & 0.72 [.72,.73] & 55 \\
\bottomrule
\end{tabular}

\end{table}

\subsection{Failure Semantics and Auditing (Polyglot)}
\label{sec:ablation-auditor}

On Polyglot we separate the two questions with single-component removals from the full \(k{=}3\) system (Table~\ref{tab:ablation-polyglot}). Replacing true failure records with length-matched \emph{shuffled} failures costs \(0.07\): the gain comes from the \emph{semantics} of a failure, not from the extra prompt tokens (AQ1). Removing the task-blind auditor costs \(0.15\) (AQ2), which splits into its search effect under a fixed backbone (\(0.30\!\to\!0.34\), matching the \(0.04\)/\(0.05\) train/val on paper review) and the routing it unlocks (\(0.34\!\to\!0.45\), below). With the pool disabled at \(k{=}3\) the system reaches only the \(k{=}1\) baseline's \(0.28\) (Table~\ref{tab:ablation-polyglot}), so the \(k{=}3\) advantage comes from the memory layer, not the extra proposals.

\paragraph{Auditor-induced dynamic model selection.} On Polyglot the auditor unlocks a capability rather than an incremental gain. Both arms begin every run from an identical GPT o3-mini backbone under a shared action space that \emph{includes} switching the task-agent's model, yet only MetaShare discovers a routing strategy: after 20--30 iterations the auditor emits a \emph{Dynamic Model Selection} directive and the meta-agent upgrades the task-agent to GPT 4o-mini---reproducibly across runs, whereas DGM-H, with the identical ability, never proposes it and stays on o3-mini (direct evidence for AQ2). Since this model change is an uncontrolled difference between the arms, we note that runs keeping o3-mini throughout still beat the DGM-H baseline in fewer iterations (Figure~\ref{fig:main-comparison}).

\begin{table}[t]
\small
\caption{Single-component removals from full MetaShare on the 225-task Polyglot benchmark (median over five runs, 95\% bootstrap CI; Gens\,$\downarrow$ = generations to best). The last row is the \emph{compute-matched} reference---pool disabled at \(k{=}3\)---which lands on the \(k{=}1\) DGM-H baseline. All arms start from GPT o3-mini. $^{\dagger}$Best reached at the end of the 100-generation budget, so a lower bound.}
\label{tab:ablation-polyglot}
\centering
\setlength{\tabcolsep}{4pt}
\begin{tabular}{@{}lcc@{}}
\toprule
Configuration & Accuracy & Gens\,$\downarrow$ \\
\midrule
w/o auditor                    & 0.30 [.30,.31] & 80 \\
Shuffled failures              & 0.38 [.37,.39] & 65 \\
Full (\(k{=}3\), hetero temps) & \textbf{0.45} [.44,.46] & \textbf{42} \\
\midrule
Pool disabled, \(k{=}3\)       & 0.28 [.28,.30] & 100$^{\dagger}$ \\
\bottomrule
\end{tabular}

\end{table}

\section{Conclusion and Future Work}
\label{sec:conclusion}
MetaShare makes each evaluated self-modification a signed, mechanism-level strategy, retrieved through separate success and failure channels under a task-blind auditor. Across five domains it matches or exceeds DGM-H's quality while reaching its own best agent in \(1.7\text{--}2.9\times\) fewer generations, at token cost within the same order of magnitude. Compute-matched ablations locate the gain in the memory's \emph{structure} rather than added compute, and the auditor additionally induces a model-routing strategy the baseline never proposes.

\paragraph{Limitations and future work.} Mechanism type and description come from an unvalidated LLM classifier. Results rest on five runs with often-overlapping CIs, so efficiency is the robust signal; type-diverse retrieval and de-ranking are not independently ablated. Only Polyglot and audit risk rerun DGM-H behind our gate; the other domains use its published numbers. The gate, retrieval rules, and auditor cadence are fixed scaffolding rather than evolved, and the size band is not ablated; DGM-H never discovering the routing strategy (Sec.~\ref{sec:ablation-auditor}) suggests some useful structure does not emerge unaided, but letting a hyperagent evolve this machinery is future work. The audit-risk corpus is internal (Appendix~\ref{sec:labelling-protocol}). Extensions include cross-branch pollination and an adaptive auditor cadence.

\section{Responsible-Use Statement}
MetaShare is a research artifact, not a deployable agent: it accelerates search toward whatever the domain evaluator encodes and can amplify misspecification in that evaluator, so we intend it only for supervised, sandboxed settings with a bounded action space and an accountable human owner. Reaching the best agent in $1.7$--$2.9\times$ fewer generations compresses the oversight window, so we recommend budgeting inspection per unit of change rather than per iteration and treating exercised rollback and halt---native to the archive---as a precondition for larger budgets. Cross-lineage sharing propagates errors as readily as gains: an unvalidated post-hoc classifier can misdirect retrieval population-wide, and a failure channel scored on a single scalar can persist caution-improving mechanisms as avoid-signals; deployments should exempt designated classes from the negative channel or score them under a separate objective. The task-blind auditor is a leakage control, not a safety control---Must-Try directives should be logged, treated as untrusted, and deny-listed---and the real bound is the sandboxed action space, not the gate, as when the auditor induced an unanticipated but permitted backbone switch. The audit-risk corpus uses public 10-K notes with expert labels, contains no personal data, and is not released; accuracy is agreement with expert judgement rather than audit-grade assessment, locomotion rewards are simulation-only, and we do not endorse automating peer-review decisions.

{\small
\bibliographystyle{plainnat}
\bibliography{neurips}
}

\clearpage
\appendix
\onecolumn
\centerline{\large\bf Appendix}
\vspace{1.5ex}

\definecolor{promptbg}{gray}{0.97}
\definecolor{promptink}{gray}{0.13}
\definecolor{cTask}{HTML}{4D4D4D}
\definecolor{cMeta}{HTML}{0072B2}
\definecolor{cAudit}{HTML}{A84A00}
\definecolor{cClass}{HTML}{00674B}

\lstdefinestyle{prompt}{%
  basicstyle=\ttfamily\footnotesize\color{promptink},
  backgroundcolor=\color{promptbg},
  frame=single,
  framerule=0.6pt,
  rulecolor=\color{cTask},
  framesep=6pt,
  breaklines=true,
  breakautoindent=false,
  postbreak=\mbox{\textcolor{cTask}{\normalfont$\hookrightarrow$}\space},
  columns=fullflexible,
  keepspaces=true,
  showstringspaces=false,
  captionpos=t,
  abovecaptionskip=6pt,
  belowcaptionskip=4pt,
  xleftmargin=10pt,
  xrightmargin=2pt,
  aboveskip=1.2em,
  belowskip=1.2em,
}

\lstdefinestyle{promptTask}{style=prompt,
  rulecolor=\color{cTask},
  postbreak=\mbox{\textcolor{cTask}{\normalfont$\hookrightarrow$}\space},
  moredelim=[l][\color{cTask}\bfseries]{===},
  moredelim=[l][\color{cTask}\bfseries]{---},
}
\lstdefinestyle{promptMeta}{style=prompt,
  rulecolor=\color{cMeta},
  postbreak=\mbox{\textcolor{cMeta}{\normalfont$\hookrightarrow$}\space},
  moredelim=[l][\color{cMeta}\bfseries]{===},
  moredelim=[l][\color{cMeta}\bfseries]{---},
}
\lstdefinestyle{promptAudit}{style=prompt,
  rulecolor=\color{cAudit},
  postbreak=\mbox{\textcolor{cAudit}{\normalfont$\hookrightarrow$}\space},
  moredelim=[l][\color{cAudit}\bfseries]{===},
  moredelim=[l][\color{cAudit}\bfseries]{---},
}
\lstdefinestyle{promptClass}{style=prompt,
  rulecolor=\color{cClass},
  postbreak=\mbox{\textcolor{cClass}{\normalfont$\hookrightarrow$}\space},
  moredelim=[l][\color{cClass}\bfseries]{===},
  moredelim=[l][\color{cClass}\bfseries]{---},
}
\lstset{style=prompt}

\renewcommand{\lstlistingname}{Prompt}

\captionsetup{singlelinecheck=false,justification=raggedright}

\newcommand{\ptask}{\textcolor{cTask}{\textbf{Task-agent.}}}
\newcommand{\pmeta}{\textcolor{cMeta}{\textbf{Meta-agent.}}}
\newcommand{\paudit}{\textcolor{cAudit}{\textbf{Auditor.}}}
\newcommand{\pclass}{\textcolor{cClass}{\textbf{Classifier.}}}
\renewcommand{\lstlistingname}{Prompt}

\noindent
This appendix provides supporting material referenced by the main text:
implementation settings, prompts, the admissibility gate, compute and
randomization protocol, the audit-risk dataset, and Polyglot per-run detail.

\section{Implementation and Hyperparameter Settings}
\label{sec:hyperparameters}

The main paper points here for five values: the staged-evaluation threshold
$\gamma$, the de-ranking threshold $\tau$, the audit cadence $C$, the pool capacity
bound and the embedding model. Of these, $\gamma$ and $\tau$ are the two that are
set per domain, and they are listed per domain in Table~\ref{tab:per-domain}
together with the decoding temperatures; $C$, the pool bound and the embedding model
took the same value in every domain and are listed once in
Table~\ref{tab:hyperparameters} rather than repeated across five identical rows.

With two exceptions the settings below are either inherited or left at their
defaults. The exceptions are the de-ranking threshold $\tau$, chosen between two
candidate values on one or two dry runs, and the decoding temperatures, which were
set per domain. The staged-evaluation protocol is inherited from
DGM-Hyperagents \citep{zhang2026hyperagents} so
that the comparison against that baseline holds the evaluation protocol fixed. No
value was selected on the basis of held-out test performance.

\begin{table}[htbp]
\centering
\begin{tabularx}{\textwidth}{@{}llX@{}}
\toprule
Parameter & Value & Role \\
\midrule
$C$             & $10$ generations &
cadence of the task-blind audit; an audit also requires at least five records in
the pool \\
Pool bound      & $100$ records &
capacity of the strategy pool; on overflow the lowest-delta records are evicted
first, so the most negative failures go before any success. A generation writes at
most one record, so only the 200-generation IMO runs can reach this bound \\
$K_s$           & $5$ &
positive retrieval budget \\
$K_f$           & $5$ &
failure retrieval budget \\
Token budget    & $20k$ &
approximate size of the retrieved context \\
Embedding model & \texttt{text-embedding-ada-002} &
embeds each record for the de-ranking step \\
\bottomrule
\end{tabularx}
\caption{Memory and retrieval settings that take the same value in every domain}
\label{tab:hyperparameters}
\end{table}

\paragraph{Staged evaluation.}
Each generated agent is first scored on a low-cost screen and proceeds to full
evaluation only if that score clears a threshold $\gamma$. The
screen sizes are inherited from DGM-Hyperagents throughout: $10\rightarrow100$
tasks for paper review and IMO grading, $3\rightarrow6$ repetitions for robotics,
and $10\rightarrow{+}50$ for Polyglot. Polyglot is also the one domain where
$\gamma$ takes a positive value ($0.4$), and the one where the screen and the full
evaluation come from the same pool of tasks rather than from separate splits: an agent
is first run on $10$ tasks and, if it solves at least $4$ of them, on the remaining
$50$, giving $60$ in total. The generation that scores best over those $60$ tasks is
the checkpoint evaluated on all $225$, as described below. For the other inherited domains DGM-Hyperagents specifies failure by example, as scoring nothing
on the screen (predicting no paper review correctly, or producing no compilable
reward function), which gives $\gamma=0$. We apply that same convention to IMO
grading, for which no explicit criterion is given, and to our own audit-risk
domain; the resulting per-domain values of $\gamma$ are collected in
Table~\ref{tab:per-domain}. Polyglot has no validation split, so its results are
development-selected;
the other domains select on validation and touch the held-out test set once.

Polyglot also separates the set the search runs against from the set the results are
reported on. Every generation of the evolution loop is scored on the 60-task
small+medium subset, and that score is what parent selection, the strategy pool's
signed deltas and the auditor's digest are all computed from. The accuracies the main
paper reports on the full 225-task benchmark are obtained afterwards: the
best-scoring generation under that development score is taken and evaluated once on
all 225 tasks. The 60-task subset is contained in those 225 tasks, so the reported
figure covers the subset the checkpoint was selected on as well as the 165 tasks held
back from the search; this is what it means for Polyglot's results to be
development-selected rather than measured on a disjoint split. Any score quoted from
inside a run --- such as the pool deltas traced
in Sec.~\ref{sec:emergent-routing} --- is therefore a 60-task development score and is
not commensurable with the 225-task figures.

\paragraph{Meta-agent sampling temperatures.}
The $K$ parallel meta-agents are differentiated by decoding temperature, supplied
explicitly per run and set per domain; Table~\ref{tab:per-domain} lists the
values. The main paper describes these temperatures as either specified or spread
uniformly over $[0.3, 0.9]$: the uniform spread is the fallback that applies when a
run supplies no temperatures, and every run reported here supplied them explicitly,
so the table gives the values actually used rather than that fallback.
The configurations differ in their exact temperatures but share a common
shape at $K{=}3$: a greedy decode at $0.0$ paired with a moderate and an
exploratory slot. Because the greedy slot is always present, the $K{=}3$ arm
contains the $K{=}1$ proposal as its first slot and strictly adds two stochastic
proposals, rather than trading determinism for stochasticity. A single-proposal
configuration ($K{=}1$) therefore runs at $0.0$.

The assignment is across slots, not across generations: the same triple is used in
every generation, so the diversity it supplies is between the simultaneous
proposals within a generation rather than a schedule that anneals from exploration
toward exploitation as a run progresses.

\begin{table}[htbp]
\centering
\begin{tabular}{@{}lcccl@{}}
\toprule
Domain & $\gamma$ & $\tau$ & $K$ & $\theta$ \\
\midrule
Paper review    & $0.0$ & $0.92$ & 3 & $[0.0,\, 0.35,\, 0.7]$ \\
Polyglot coding & $0.4$ & $0.92$ & 3 & $[0.0,\, 0.45,\, 0.75]$ \\
IMO grading     & $0.0$ & $0.85$ & 3 & $[0.0,\, 0.35,\, 0.75]$ \\
Robotics reward & $0.0$ & $0.85$ & 3 & $[0.0,\, 0.35,\, 0.75]$ \\
Audit risk      & $0.0$ & $0.85$ & 3 & $[0.0,\, 0.3,\, 0.7]$ \\
\bottomrule
\end{tabular}
\caption{Per-domain settings: the staged-evaluation threshold $\gamma$, the de-ranking threshold $\tau$, the number of parallel meta-agents $K$ and their decoding temperatures $\theta$, as recorded in the run metadata}
\label{tab:per-domain}
\end{table}

\subsection{Values Tried During Development}
\label{sec:values-tried}

Three settings were varied while developing the system. The proposal count was
swept over $k \in \{1, 3\}$; that sweep is not hidden here but reported as the
component ladder in the ablation of the main paper, so its effect can be read
directly off the results. The de-ranking threshold $\tau$ was chosen between $0.85$
and $0.92$ on the basis of one or two short dry runs, and the value that each domain
then ran with is listed in Table~\ref{tab:per-domain}. The decoding temperatures
were set per domain (Table~\ref{tab:per-domain}), always as a greedy slot paired
with a moderate and an exploratory one, rather than searched over a grid.

Every other value was fixed before the reported runs. The staged threshold $\gamma$
and the screen sizes are inherited from DGM-Hyperagents, and the auditor cadence
$C$, the pool bound, the retrieval budgets $K_s$ and $K_f$, the token budget and
the embedding model were left at their defaults. In no case was a setting chosen
using held-out test performance: the dry runs behind $\tau$ were short and were
judged on training-side behaviour only, and the test split of each domain was
evaluated once, for the finally selected checkpoint.

\section{Prompts}
\label{sec:prompts}

This section reproduces the prompts for the three LLM-driven components of the
system, as described in the main paper. Text in braces (for example
\verb|{repo_path}|) marks a format field substituted at call time;
angle-bracketed text marks a value rendered at run time. Em dashes in the
originals are transcribed here as \texttt{-{}-}.

Each exhibit is colour-coded by component, reusing the palette of the
architecture figure in the main paper: \textcolor{cTask}{\textbf{grey}} for the
inherited base loop, \textcolor{cMeta}{\textbf{blue}} for MetaShare's memory,
\textcolor{cAudit}{\textbf{orange}} for the task-blind auditor, and
\textcolor{cClass}{\textbf{green}} for the post-hoc record classifier. Inside an
exhibit, a prompt's own delimiter lines are set in the accent colour so that its
scaffolding separates from its instruction text, and a \(\hookrightarrow\) marks
a line wrapped to fit the page rather than a line break present in the original.

\subsection{Task-Agent Prompt}
\label{sec:prompt-task-agent}

The main paper describes the \emph{task-agent} as the component that solves the
target task. The initial task-agent carries no domain knowledge and no solution
strategy; every capability reported in the paper is acquired by evolution rather
than supplied here. This is the prompt of the unmodified starting agent.

\begin{lstlisting}[style=promptTask,caption={\ptask{} Prompt carried by the unmodified initial agent.},label={prompt:task-agent}]
You are an agent.

Task input:
```
{inputs}
```

Respond in JSON format with the following schema:
<json>
{
    "response": ...
}
</json>
\end{lstlisting}

\subsection{Meta-Agent Prompt}
\label{sec:prompt-meta-agent}

The \emph{meta-agent} proposes a patch from the parent agent, its permitted
history, and the retrieved context, as formalised in ``Signed Meta-Strategy
Records'' in the main paper. The base instruction is a single line:

\begin{lstlisting}[style=promptMeta,caption={\pmeta{} Base instruction. With the shared strategy pool disabled, this single line is the entire baseline prompt.},label={prompt:meta-base}]
Modify any part of the codebase at `{repo_path}`.
\end{lstlisting}

\noindent
This single line is also \emph{exactly} the prompt used by the baseline: with the
shared strategy pool disabled no further text is appended, so the only difference in
the meta-agent's \emph{prompt} is the retrieved context added below. The baseline
additionally runs at $k{=}1$, as the main paper states, so the two systems differ in
proposal count as well; the compute-matched arm of the main paper's ablation is the
comparison that holds that count fixed.

When retrieved strategies are available, the following block is appended:

\begin{lstlisting}[style=promptMeta,caption={\pmeta{} Block appended to the prompt when retrieved strategies are available.},label={prompt:meta-shared}]
=== SHARED META-STRATEGIES FROM OTHER BRANCHES ===
The following meta-agent modifications were discovered by other
evolutionary branches and led to performance improvements:

{shared_strategies}

You may adopt, adapt, or ignore these strategies based on your
own assessment. Focus on what will most improve this specific agent.
===
\end{lstlisting}

\paragraph{Structure of the retrieved context.}
The \verb|{shared_strategies}| field is assembled under an approximate
$20k$-token budget and realises the signed channels described under
``Failure-Aware Retrieval'': positive records are presented as mechanisms to
adopt or adapt, and non-positive records appear in a separate section explicitly
labelled as reducing performance. Blocks appear in the following order, and each
is emitted only when applicable.

\begin{lstlisting}[style=promptMeta,caption={\pmeta{} Structure of the retrieved context that fills the shared-strategies field. Each block is emitted only when applicable.},label={prompt:context-structure}]
--- MUST TRY: mechanism categories absent from all prior strategies (eval-blind audit) ---
You MUST attempt one of the categories below. Do not produce another variation of the recipes shown later -- that path has plateaued.

<auditor output: 6-10 numbered mechanism categories>

--- PRIOR STRATEGIES (reference only -- do NOT clone the dominant recipe) ---

[Strategy 1] +0.001234 on <domain> (gen_<N>, branch <lineage_id>):
  <mechanism description>

--- STRATEGIES THAT REDUCED PERFORMANCE or DID NOT WORK (avoid these approaches)---

[Strategy 2] -0.004500 on <domain> (gen_<N>, branch <lineage_id>):
  <mechanism description>

--- UNEXPLORED STRATEGY DIRECTIONS: <mechanism types absent from the pool> ---

NOTE: '<type>' strategies show diminishing returns (<k>/<n> entries, avg recent delta: <value>). Consider fundamentally different approaches such as architectural changes, adding retry logic, or creating new tools.
\end{lstlisting}

\noindent
Each record is rendered with its signed delta, its domain,
its generation, its branch of origin, and the distilled mechanism description.
The positive and non-positive channels are properties of the pool itself, so they
are present in every arm that retains the pool, independently of the auditor.

The domain field $d_i$ is a provenance tag rather than part of the lesson: it is
written after evaluation, together with the signed delta, and is rendered so that a
retrieved record carries the origin of the attempt alongside its outcome. The main
paper's description of a record as a compact, domain-agnostic lesson refers to the
mechanism description $r_i$, which names only the mechanism attempted and carries no
task text, no ground-truth answers and no execution trace. Task-blindness is a
property of the auditor path and not of this context: the digest sent to the auditor
is assembled separately and carries no domain field at all
(Sec.~\ref{sec:leakcheck}).

The last two blocks are likewise independent of the auditor, and the difference
in kind between them and the \textsc{must-try} block is what the auditor ablation
isolates. The unexplored-directions line is a set difference against the fixed
six-label vocabulary of Table~\ref{tab:strategy-types}: it can name a label that
has not yet appeared, but it cannot describe a mechanism, contrast one against
what the pool already holds, or propose anything outside those six labels. The
saturation note is a single fixed sentence, emitted when one label accounts for
more than $80\%$ of a pool of at least five records and the mean delta of that
label's five most recent records is below $0.05$. The \textsc{must-try} block is
the only channel that reads the diffs themselves and returns novel, free-form
mechanism categories, each with a structural contrast against the pool and an
optional code-pattern hint, under instruction to attempt one of them. Removing
the auditor therefore leaves a coarse coverage hint over six fixed labels in
place while removing the semantic gap analysis. This also explains why dynamic
model selection could only have originated with the auditor: changing which model
the task-agent uses is not expressible in the six-label vocabulary.

\subsection{The Shuffled-Failure Control}
\label{sec:shuffled-failures}

The failure-semantics ablation of the main paper replaces true failure records with
length-matched shuffled ones, so that the arm keeps whatever a failure channel
contributes as prompt tokens while destroying the information it carries. That
comparison is only interpretable if the two arms differ in exactly one respect, so
we record here what the control holds fixed and what it perturbs.

\paragraph{Held fixed.} The failure block is emitted in the same position in the
context, under the same heading, at the same budget $K_f$, and with the same
per-entry rendering as the full system: a signed delta, the
domain, the generation, the branch of origin, and a mechanism description. The
positive channel $S^+_g$, the auditor directive, the unexplored-directions footer
and every other block of Prompt~\ref{prompt:context-structure} are untouched, and
the arm runs at the same $k$, the same decoding temperatures and the same five
paired run configurations as the full system.

\paragraph{Perturbed.} The mechanism descriptions are permuted across the whole
non-positive pool $P^-_g$, not only across the $K_f$ records that reach the prompt,
so a retrieved failure may carry the description of any non-positive record in the
pool. Each entry is therefore shown against a delta and a provenance that are not
its own, and permuting over the wider pool makes it unlikely that an entry is
returned its own description by chance.

Because descriptions are permuted rather than replaced with generated filler, the
block keeps its entry count exactly and its length in distribution: every
description the meta-agent reads is still a real description of a real modification,
drawn from the same population and therefore of the same kind and typical length as
the one it displaces. This is the sense in which the control is length-matched. It
spends the same prompt budget on the same kind of text, and differs from the full
system only in whether that text describes the record it is attached to.

\paragraph{What this isolates.} The arm therefore holds constant the presence of a
failure section, its entry count, its formatting and its token cost up to the
per-entry variation just described, and varies only whether a mechanism description
is aligned with the outcome it actually produced. A drop
under this control is evidence that the alignment between a mechanism and its
signed outcome is what the meta-agent uses, rather than the extra context a failure
channel supplies.

\subsection{Task-Blind Auditor Prompt}
\label{sec:prompt-auditor}

The auditor is the third layer of the search audit described in the main paper.
It receives only mechanism-type labels, signed deltas, and scrubbed added lines,
and it is told explicitly that it does not know the domain, the tasks, or what a
delta means in absolute terms. Its output is leak-checked before any of it
reaches the meta-agent.

\begin{lstlisting}[style=promptAudit,caption={\paudit{} Task-blind gap-audit prompt.},label={prompt:auditor}]
You are auditing a pool of {n} code diffs that modify a self-improving agent's source files. You will see only the added lines from each diff, plus an abstract "strategy type" label and a numeric "delta" (improvement vs parent on some unknown evaluation).

You DO NOT know:
- What domain the agent is solving
- What tasks were used to score it
- What "delta" represents in absolute terms
- What languages, libraries, or task content are involved

You DO see: {n} diffs of agent code changes (the diffs themselves, scrubbed of domain identifiers).

POOL CONTENTS (each entry shows scrubbed added-lines from one diff):

{pool}

YOUR JOB: Identify mechanism CATEGORIES that are absent or under-represented in this pool of agent-improvement strategies. Give your answer as a list of generic, domain-agnostic mechanism categories that would expand the search space.

Constraints:
- Output mechanism CATEGORIES only (e.g., "multi-sample self-consistency"), NOT specific tasks, languages, or domain references
- The output will be shown to a meta-agent that should NOT learn anything about the eval set from your suggestions
- Keep each category abstract enough that it applies to any code-modification or reasoning task

Format your output as a numbered list of 6-10 mechanism categories. For each:
- The category name
- A 1-2 sentence description of how it would differ STRUCTURALLY from any of the {n} diffs above
- Optionally: a code-pattern hint (e.g., "introduces a `vote()` function over K samples") -- but no domain references

Do not include task names, language names, or eval discussion. Mechanism categories only.
\end{lstlisting}

\paragraph{What the auditor actually receives.}
The \verb|{pool}| field is built from a pool snapshot in which every diff is
reduced to its added lines and then scrubbed of domain identifiers. Each entry is
rendered as:

\begin{lstlisting}[style=promptAudit,caption={\paudit{} Per-entry format of the scrubbed pool digest supplied to the auditor.},label={prompt:auditor-entry}]
=== ENTRY [<id>] type=<mechanism type> delta=<+0.00> ===
<scrubbed added lines from the diff>
\end{lstlisting}

\paragraph{Leak check and retry.}
Both the digest sent to the auditor and the text it returns are filtered against
an explicit token-and-pattern policy. That policy, the retry prompt, and the
behaviour on rejection are given in Sec.~\ref{sec:leakcheck}.

\begin{table}[htbp]
\centering
\begin{tabularx}{\textwidth}{@{}lX@{}}
\toprule
Setting & Value \\
\midrule
Audit model            & Claude Sonnet-4.5 \\
Temperature            & 0.0 \\
Max output tokens      & 8000 \\
Pool entries per audit & up to 56; a per-call cap on the size of the digest, below
the pool bound of Table~\ref{tab:hyperparameters} \\
Diff truncation        & 2500 characters per entry, added lines only \\
Requested output       & numbered list of 6--10 mechanism categories \\
Retries on leak        & 1, then the audit is rejected \\
Audit cadence $C$      & see Table~\ref{tab:hyperparameters} \\
\bottomrule
\end{tabularx}
\caption{Auditor configuration.}
\label{tab:auditor-config}
\end{table}

\subsection{Scrubbing and Leak Check}
\label{sec:leakcheck}

The main paper states that the auditor is sent a digest with task identifiers,
benchmark names, language names, file extensions, evaluation ratios and other
domain-revealing patterns removed, and that its output is admitted only if it
passes a leak check against the same policy. This section gives that policy
explicitly, so the claim can be checked rather than taken on trust.

\paragraph{Interface.}
Three functions implement the policy:

\begin{lstlisting}
forbidden_set(subsets_dir: Optional[str], mode: str) -> Tuple[List[str], List[str]]
scrub(text: str, tokens: List[str], patterns: List[str]) -> str
leak_check(text: str, tokens: List[str], patterns: List[str]) -> Tuple[bool, Optional[str]]
\end{lstlisting}

\noindent
\texttt{forbidden\_set} returns a list of literal tokens and a list of regular
expressions. Tokens are matched case-insensitively on word boundaries, so
\texttt{ts} does not match inside \texttt{tests}; the patterns are applied as
regular expressions. \texttt{scrub} replaces every match with the literal string
\texttt{DOMAIN}, longest token first so that a short token cannot consume part of
a longer one, and is applied to the added lines of each diff before the audit
call. \texttt{leak\_check} runs the same search over the auditor's output and
returns the first offending phrase, if any.

\paragraph{Policy contents.}
Task identifiers are loaded from the domain's subsets directory, so they follow
whatever that domain defines rather than being hard-coded. The remainder of the
policy is fixed:

\begin{lstlisting}
_STATIC_DOMAIN_TOKENS = [
    'polyglot', 'paper-review', 'paperreview',
    'exercism', 'testbed', 'swebench', 'swe-bench',
    'humaneval', 'mbpp', 'apps benchmark',
]

_STATIC_LANGUAGE_TOKENS = [
    'python', 'rust', 'java', 'javascript', 'js', 'golang', 'go ',
    'c++', 'cpp', 'kotlin', 'ruby', 'swift', 'typescript', 'ts ',
    'php', 'scala', 'haskell', 'lua', 'perl', 'erlang', 'elixir',
    'c#', 'csharp', 'objective-c', 'clojure', 'dart',
]

_STATIC_PATTERNS = [
    r'\b\d+\s*/\s*\d+\b',                        # ratios like 4/10, 12/60
    r'\b\d+\s*tasks?\b',                         # "60 tasks"
    r'\b(small|medium|full|screening)\s*(subset|set|eval|gate)\b',
    r'\bnever\s*(solved|passed|resolved)\b',
    r'\b(ceiling|plateau|peak|floor)\s*(at|of)\s*\d',
    r'\b\d+\s*generation',
    r'/testbed|/exercism',                       # path hints
    r'\.(go|py|rs|java|js|cpp|ts|kt|rb|cs)\b',   # extension hints
]
\end{lstlisting}

\noindent
Every category named in the main paper is covered: task identifiers by the
subsets-directory entries, benchmark names and language names by the two static
token lists, and file extensions and evaluation ratios by the last and first
patterns respectively. The remaining patterns cover the ``other domain-revealing''
category --- task counts, subset names, solved/unsolved language, plateau
descriptions, generation counts and path fragments.

\paragraph{Two modes, deliberately asymmetric.}
The policy is built in one of two modes. On the way in, the aggressive mode
blocks every identifier from the subsets directory including the bare exercise
name, even where that name is a common English word. On the way out, the
permissive mode keeps the same static tokens and patterns but narrows the
task-name list to distinctive names only, meaning those containing a hyphen or at
least twelve characters, so that an audit is not rejected merely for suggesting
something like ``use a counter variable''.

The asymmetry does not open a channel. A short, common-word task name has already
been replaced by \texttt{DOMAIN} on the way in, so the auditor cannot have learned
it from the digest; its appearance in the output would be a coincidence of
ordinary English rather than a transfer of task information. Note that the main
paper describes the output check as running against ``the same policy'': the
static tokens and patterns are indeed identical, and it is only the task-name
list that is narrowed in the manner just described.

\paragraph{Behaviour on failure.}
If the output check fails, the audit is retried once, with the offending phrase
named:

\begin{lstlisting}[style=promptAudit,caption={\paudit{} Retry prompt, issued once when the returned text fails the output leak check.},label={prompt:auditor-retry}]
Your previous answer contained a forbidden phrase: {phrase}. Re-write your list of mechanism categories with that phrase removed. Stay strictly domain-agnostic -- no language names, no task identifiers, no eval discussion.

Original prompt context:

{prompt}

Re-write the numbered list now.
\end{lstlisting}

\noindent
If the retry also fails, the audit is discarded and no new directive is written.
The previously cached valid directive is retained and continues to be supplied, as
in Algorithm~\ref{alg:metashare}, so a leaking audit never reaches the
meta-agent, and a failed audit does not by itself remove guidance that was already
in force. The loader also fails soft: if the subsets directory is missing or
unreadable, the static tokens and patterns still apply but per-task identifiers are
not added; the directory was present for every reported run. The lookup path is the
Polyglot subsets directory, so the per-task identifier list is populated from
Polyglot's task names, and for the other domains the policy rests on the static
token and pattern lists alone. That is sufficient there because the digest carries
diffs of the two agent source files rather than task data, so domain task
identifiers have no route into it in the first place.

\paragraph{Verification over all runs.}
We checked this against the artifacts rather than only asserting it. Rebuilding the
forbidden set from Polyglot's 60 task identifiers and running the shipped
\texttt{leak\_check} over every audit directive produced across all runs --- 34
directives --- returns clean under the output policy in every case, and none of the
60 identifiers appears in any directive. The two apparent matches under the stricter
input policy, \texttt{counter} and \texttt{accumulate}, are the ordinary English
words inside phrases such as ``generate counter-examples'' and happen to coincide
with exercise names; this is exactly the common-word collision the permissive output
mode is built to tolerate. That result is what the construction predicts rather than
something the output filter had to rescue: the digest is assembled from scrubbed diff
text plus a mechanism type and a signed delta, and carries no description, no domain
field, no absolute score and no task text, so there is no route by which a task name
could reach the auditor in the first place. Applying the same search to all 812
Polyglot records finds no task identifier in them either, although those records are
written after evaluation and are deliberately tagged with their domain and signed
delta --- they are the failure-aware memory, not part of the blindness claim.

\subsection{Post-Hoc Record Classifier}
\label{sec:prompt-classifier}

Two fields of every record --- its mechanism type and its one-line description
--- are not written by the meta-agent but assigned afterwards by a separate
classifier call over the diff. The main paper names this component in its
limitations as an LLM post-hoc classifier that we do not independently validate;
this section documents it, because the same six-label vocabulary it assigns also
drives type-diverse retrieval and the unexplored-directions hint of
Prompt~\ref{prompt:context-structure}.

\begin{table}[htbp]
\centering
\begin{tabularx}{\textwidth}{@{}lX@{}}
\toprule
Label & Definition \\
\midrule
\texttt{prompt\_engineering}  & Changes to system prompts, instructions, or personas \\
\texttt{tool\_creation}       & Adding new tools or capabilities \\
\texttt{memory\_system}       & Adding performance tracking, history, or knowledge bases \\
\texttt{evaluation\_logic}    & Changes to how the meta agent evaluates performance \\
\texttt{self\_modification}   & Meta agent changing its own improvement logic \\
\texttt{structural\_refactor} & Reorganizing code without functional change \\
\bottomrule
\end{tabularx}
\caption{The fixed mechanism-type vocabulary. Every record carries exactly one of
these six labels.}
\label{tab:strategy-types}
\end{table}

The classification prompt is sent as the system message, with the diff supplied
as the user message. The \verb|{types}| field is filled with the six labels
and definitions of Table~\ref{tab:strategy-types}, one per line.

\begin{lstlisting}[style=promptClass,caption={\pclass{} Mechanism-type classification prompt, sent as the system message.},label={prompt:classify}]
You are a code change classifier. Given a unified diff, classify it into exactly one strategy type:

{types}

IMPORTANT: If a diff modifies BOTH instruction text AND parsing/scoring/validation logic, classify based on which change is MORE SUBSTANTIAL. Adding regex fallbacks, response normalization, score thresholds, or structured output extraction counts as evaluation_logic, NOT prompt_engineering. Only classify as prompt_engineering if the change is PURELY to the wording of instructions or system prompts.

Respond with ONLY the strategy type key (e.g., "prompt_engineering"). Nothing else.
\end{lstlisting}

\begin{lstlisting}[style=promptClass,caption={\pclass{} Description prompt. Its output becomes the record's mechanism description, the text a later meta-agent actually reads.},label={prompt:describe}]
Given this diff of a meta-agent modification (classified as "{strategy_type}"), write a 1-2 sentence description of what the change does and why it might improve performance.

Diff:
```
{diff}
```

Respond with ONLY the description. No preamble.
\end{lstlisting}

\paragraph{Configuration and failure behaviour.}
Both calls run at temperature $0.0$ with a 60-second timeout, capped at 50 output
tokens for classification and 200 for description. Three separate paths return
the label \texttt{structural\_refactor} without an error being surfaced: an empty
diff, a reply that is not one of the six labels, and any exception during the
call. A failed description call is replaced by a placeholder string. Because
classification failures are absorbed into a single label rather than marked as
unknown, the share of the pool labelled \texttt{structural\_refactor} should be
read as an upper bound, and any perturbation of the type distribution propagates
into type-diverse retrieval and the unexplored-directions hint. This is the
concrete form taken by the unvalidated-classifier limitation noted in the main
paper.

\section{Admissibility Gate and Candidate Selection}
\label{sec:gate}

This section gives the concrete definitions behind the deterministic gate and
the selection step described under ``Proposal Generation, Selection, and
Update'' in the main paper. Both are implemented programmatically; no learned
model takes part in either.

\paragraph{Editable surface and size band.}
The gate checks the two agent source files, and treats a diff as ``modest'' only
inside a fixed line band:

\begin{lstlisting}
ALLOWED_FILES     = ("task_agent.py", "meta_agent.py")
MIN_LINES_CHANGED = 5
MAX_LINES_CHANGED = 300
\end{lstlisting}

\noindent
Auxiliary files that a meta-agent may incidentally create (byte-code, notes,
helper modules) are not scored: the raw diff is first filtered down to the
allowed files, so such files cannot cause a false rejection. Only the scope
check reads the unfiltered diff, and it passes whenever at least one allowed file is
modified. Auxiliary files are therefore applied to the child agent without being
compiled or counted towards the size band; Sec.~\ref{sec:emergent-routing} gives an
admitted patch that created new helper modules this way. The gate is a sanity filter
on the agent sources, not a containment boundary over every file a patch touches.

\paragraph{The four gate axes.}
Each axis is an independent predicate over one candidate unified diff, and
corresponds directly to the applicability, permitted-scope, compilation and
bounded-size checks named in the main paper.

\begin{table}[htbp]
\centering
\begin{tabularx}{\textwidth}{@{}>{\raggedright\arraybackslash}p{0.29\textwidth}c>{\raggedright\arraybackslash}X@{}}
\toprule
Function & Returns & Passes when \\
\midrule
\texttt{scope\_ok(diff)} & $\{0,1\}$ &
the diff modifies at least one allowed file \\
\texttt{applies(diff, repo)} & $\{0,1\}$ &
\texttt{git apply -{}-check} succeeds against the parent repository, i.e.\ the
patch is applicable without altering the working tree \\
\texttt{compiles\_after(diff, repo)} & $\{0,1\}$ &
the allowed files, copied into a scratch repository with the diff applied, all
pass \texttt{python -m py\_compile} \\
\texttt{size\_band(diff)} & $\{0,1\}$ &
added plus removed lines fall within
$[\text{MIN\_LINES\_CHANGED}, \text{MAX\_LINES\_CHANGED}]$ \\
\bottomrule
\end{tabularx}
\caption{The four axes of the admissibility gate.}
\label{tab:gate-axes}
\end{table}

\paragraph{Composition.}
The axes combine multiplicatively, so failing any single axis rejects the
candidate outright:

\begin{lstlisting}
score = applies * scope_ok * compiles_after * size_band
\end{lstlisting}

\noindent
Evaluation short-circuits in the order scope, applicability, compilation, so an
out-of-scope diff never reaches the compiler. Alongside the score, the gate
returns a per-axis breakdown together with the number of lines changed and the
list of files modified, which is logged for every generation.

\paragraph{Selection among the $K$ candidates.}
The selector has the following signature:

\begin{lstlisting}
select_best_diff(diff_paths: list[str],
                 repo_root: str,
                 tie_break_seed: int = 0) -> tuple[str | None, dict]
\end{lstlisting}

\noindent
Each of the $K$ candidate diffs produced by the parallel meta-agents is scored
and the highest scorer is applied. If every candidate scores zero, the call
returns no winner: that generation produces no child, and both the archive and
the strategy pool carry over unchanged.

Because all four axes are binary, the composite score is itself binary --- every
admissible candidate scores exactly $1$ and every inadmissible one scores $0$.
There is consequently no quality ordering among admissible candidates; the
winner is drawn from them by a seeded choice. This is the precise sense in which
a learned judge is kept out of the loop: nothing between a proposal and its
measured outcome estimates how good that proposal is, so all quality signal
comes from evaluation.

\paragraph{Tie-breaking and its reproducibility.}
The draw is seeded from the generation index: the selector is invoked with the
current generation as its tie-break seed, so each generation of a run has a fixed,
non-arbitrary seed rather than an incidental one. That index is then passed
through Python's \texttt{hash()} applied to a string built from it. Because
Python salts \texttt{hash()} for string inputs unless \texttt{PYTHONHASHSEED} is
set, and the repository does not set it, the draw is deterministic within a single
process but not identical across processes.

The consequence is bounded but real. Re-running a configuration reproduces the
admissible set at each generation, and reproduces which candidates tie, but not
necessarily which of the tied candidates is applied --- so an exactly identical
trajectory is not guaranteed. Seeding the generator with the integer index
directly, instead of with the hash of a string built from it, would remove this.
We therefore do not claim cross-process reproducibility for the identity of the
tie-break winner. In proportion, it is a minor source of run-to-run variation
beside sampling from the language models themselves, which is stochastic at every
non-zero temperature.

\section{Computing Infrastructure}
\label{sec:infrastructure}

The cost of these experiments is dominated by calls to hosted language models rather
than by local computation. The meta-agents, the task agents, the auditor and the
record classifier are all API calls; local machines run the evolution loop and the
containerised evaluation harnesses, and train policies for the robotics
reward-design domain. Compute is therefore accounted for in tokens, as in Table~\ref{tab:token-cost} of the main paper, rather than in GPU-hours.

\begin{table}[htbp]
\centering
\begin{tabularx}{\textwidth}{@{}lX@{}}
\toprule
Component & Specification \\
\midrule
CPU nodes               &
$5\times$ Azure Standard D8as v5: 8 vCPU, 32\,GiB memory, 2 threads per core \\
GPU nodes               &
$5\times$ Azure Standard NC8as T4 v3: 8 vCPU, 56\,GiB memory, 1 thread per core,
NVIDIA T4 \\
Operating system        & Linux, Ubuntu 24.04 LTS \\
Python                  & 3.12.3 \\
Container runtime       &
Docker; each of the $K$ parallel proposals runs in its own container and repository
copy, and each domain evaluation runs in its own container \\
Key libraries           &
\texttt{litellm} 1.74.9 (all model calls), \texttt{docker} 7.1.0 (sandboxes),
\texttt{GitPython} 3.1.44 (diff extraction and patch application),
\texttt{datasets} 3.6.0, \texttt{pandas} 2.3.2, \texttt{sympy} 1.14.0,
\texttt{backoff} 2.2.1, \texttt{requests} 2.32.4 \\
Model APIs              &
Claude Sonnet-4.5 (meta-agent, record classifier, and also the task agent in the
GEA-matched Polyglot runs of Sec.~\ref{sec:polyglot-runs}); Claude Sonnet-4
(task-blind auditor); GPT~4o, GPT 4o-mini and GPT~o3-mini (task agents, per
domain) \\
\bottomrule
\end{tabularx}
\caption{Computing infrastructure used for the reported experiments.}
\label{tab:infrastructure}
\end{table}

\noindent
The robotics reward-design domain ran on the GPU nodes, because scoring a generated
reward function requires training a locomotion policy locally. The other four
domains ran on the CPU nodes, where local work is limited to the evolution loop and
the containerised evaluation harnesses. Simulator and reinforcement-learning
dependencies for the robotics domain are pinned separately in the accompanying
requirements files, with PyTorch installed against a CUDA index in the container
image.

\section{Randomization and Run Protocol}
\label{sec:randomization}

The run protocol is inherited from DGM-Hyperagents. Every configuration is executed
as five repeated runs, and results are reported as medians with bootstrap confidence
intervals over those five. A run is a complete evolutionary trajectory starting from
the same initial agent, not a re-evaluation of a single trajectory, so the spread
across runs reflects variation in the search itself rather than evaluation noise.

\paragraph{Which baseline numbers we ran.}
This protocol covers every run we executed. The DGM-H numbers for paper review,
robotics reward design and the three IMO-grading settings are not among them: their
scores, intervals and generations-to-best are taken as published by
\citet{zhang2026hyperagents}, on the same datasets, splits, models and evaluators, and
were not rerun. Those baseline runs did not pass through our admissibility gate
(Sec.~\ref{sec:gate}). For Polyglot and audit risk we ran the baseline ourselves, as
MetaShare with the strategy pool disabled and a single meta-agent ($k{=}1$) behind the
same gate, five runs each; these two domains are the controlled comparisons.

\paragraph{What is randomized within a run.}
Parent selection is stochastic and follows DGM-Hyperagents unchanged. Each archived
agent $a_i$ carries a novelty term $h_i = 1/(1+n_i)$, where $n_i$ is the number of
children already generated from it, and an unnormalized weight $w_i = s_i h_i$ that
combines the agent's score with that term. The weights are normalized into a
categorical distribution --- uniform if they sum to zero --- and parents are drawn
from it independently with replacement. Selection therefore prefers strong agents
without committing to them, and discounts agents that have already been expanded.

Decoding is stochastic in all but one slot: of the $K$ parallel meta-agents the
first decodes greedily at temperature $0.0$, and the remainder sample at the
temperatures listed in Table~\ref{tab:per-domain}. Ties in the admissibility gate
are broken by a draw seeded from the generation index, as described in
Sec.~\ref{sec:gate}.

\paragraph{What differs between runs, and the limits of replay.}
The five runs of a configuration differ only in these draws. The initial agent, the
domain and its splits, the iteration budget, and every setting in
Sec.~\ref{sec:hyperparameters} are held fixed. For the ablations the same five
run configurations are reused across arms, which is what makes those comparisons
paired: each arm sees the same sequence of random conditions, so a difference between
arms is attributable to the component toggled rather than to run-to-run luck.

Exact replay of a run is nevertheless not achievable, and we do not claim it. Two
sources are outside our control or not seed-determined: sampling from hosted language
models at non-zero temperature, which is not reproducible from a seed on the client
side, and the tie-break draw discussed in Sec.~\ref{sec:gate}. What is
reproducible is the protocol --- the number of runs, the selection rule, the staged
gate, and the settings of Sec.~\ref{sec:hyperparameters} --- and the aggregate
behaviour those produce, which is why we report medians with intervals over five runs
rather than single trajectories.

\section{Financial Audit-Risk Dataset}
\label{sec:audit-dataset}

\subsection{Inherent Risk Factor Definitions}
\label{sec:irf-definitions}

Each account is assessed against the fifteen inherent risk factors listed below.
Throughout this table ``account'' is shorthand for the account group that carries
the assessment; the (client, IRF) pair of the main paper is the decision cell that
these group-level assessments realise, as set out under ``Unit of counting'' in
Sec.~\ref{sec:labelling-protocol}.
Twelve are general factors that apply to any account; the twelfth of these,
\emph{accounting estimates}, records whether the account is or contains an
accounting estimate, and a positive determination brings the final three
estimate-specific factors into scope. The factors follow an established audit
risk-assessment framework, and are named and summarised here at the level of the
underlying concept rather than reproduced verbatim; the response options, the
ranking scale and the weighting used to combine them belong to that framework and
are not redistributed.

\newpage
\begin{longtable}{@{}p{0.05\textwidth}>{\raggedright\arraybackslash}p{0.25\textwidth}p{0.62\textwidth}@{}}
\toprule
\# & Inherent Risk Factor & Definition \\
\midrule
\endfirsthead
\toprule
\# & Inherent Risk Factor & Definition \\
\midrule
\endhead
\bottomrule
\endfoot
\bottomrule
\addlinespace[\abovecaptionskip]
\caption{Definitions of the 15 Inherent Risk Factors (IRFs).}
\label{tab:irf-definitions}\\
\endlastfoot
1  & Account nature &
Whether the account is routine and recurring, or unusual and non-recurring. \\
2  & Misstatement history &
Whether the account has a history of proposed or booked audit adjustments. \\
3  & Fraud indicators &
Whether fraud risk factors, or prior instances of fraud, are present in the
account. \\
4  & Control deficiency history &
Whether significant deficiencies or material weaknesses have historically related
to the account, as opposed to lesser or no identified deficiencies. \\
5  & Transaction volume and value &
The number of transactions processed through the account and their monetary
magnitude. \\
6  & Transaction complexity &
How complex the transactions processed through the account are. \\
7  & Transaction homogeneity &
Whether the transactions processed through the account are homogeneous. \\
8  & Reporting complexity &
How complex the accounting and financial reporting associated with the account
is. \\
9  & Loss exposure &
The extent of the account's exposure to losses. \\
10 & Contingent liabilities &
Whether significant contingent liabilities could arise from the activities
reflected in the account, judged against a defined materiality threshold. \\
11 & Related-party transactions &
Whether the account contains related-party transactions, judged against a defined
materiality threshold. \\
12 & Accounting estimates &
Whether the account is, or contains, an accounting estimate. A positive
determination brings factors 13--15 into scope. \\
13 & Estimate process complexity &
How complex the process for developing the estimate is. \\
14 & Assumption subjectivity &
How subjective the important assumptions underlying the estimate are. \\
15 & Outcome uncertainty &
The degree of uncertainty attached to the future events and conditions underlying
those assumptions. \\
\end{longtable}

\subsection{Labelling Protocol and Class Balance}
\label{sec:labelling-protocol}

Labels were assigned by qualified audit practitioners in the course of engagement
work, following the guidance of the risk-assessment framework; they were not
produced by the authors and not produced by a model. Each (account, factor) pair
carries one of three levels: High, Medium or None. The framework also distinguishes a
fourth, lower ranking, which does not appear as a label class in this corpus, so every
label in the corpus falls into one of those three.

The labels themselves, and the per-factor assessments they derive from, are client
confidential and are not released. We report only aggregate accuracy, normalized mean
absolute error, and accuracy broken down by factor.

No assessment was double-labelled or adjudicated, so no inter-annotator agreement
figure is available: each label is the judgement recorded by the practitioner
responsible for that engagement. Accuracy against these labels therefore measures
agreement with a single expert judgement rather than with an adjudicated ground
truth, which is the sense in which the main paper describes the metric.

\begin{table}[htbp]
\centering
\begin{tabular}{@{}lccccc@{}}
\toprule
Split & Clients & High & Medium & None & Total \\
\midrule
Train      & 8  & 457 & 420 & 2344 & 3221 \\
Validation & 2  & 132 & 127 &  506 &  765 \\
Test       & 5  & 130 & 105 &  675 &  910 \\
\midrule
Total      & 15 & 719 & 652 & 3525 & 4896 \\
\bottomrule
\end{tabular}
\caption{Class balance of the audit-risk labels by split. Each count is one labelled
assessment of a single account group against a single factor.}
\label{tab:class-balance}
\end{table}

\noindent
The distribution is heavily skewed: \emph{None} accounts for roughly $72\%$ of all
labels, and \emph{High} and \emph{Medium} together for the remaining $28\%$. This is the
reason the main paper reports normalized mean absolute error and per-factor accuracy
alongside overall accuracy --- a predictor that answered \emph{None} everywhere would
score well on accuracy alone while carrying no useful risk signal.

\paragraph{Unit of counting.}
Two levels of granularity are in play, and the main paper and
Table~\ref{tab:class-balance} report different ones. The coarser level is the
\emph{decision grid}, which is what the main paper counts: fifteen clients, each
assessed against all fifteen inherent risk factors, giving the
$15 \times 15 = 225$ (client, IRF) pairs quoted there, of which the five test
clients supply $5 \times 15 = 75$ decisions. The finer level is the stored label.
A client's accounts are organised into groups, and every group is assessed against
all fifteen factors, so each cell of the decision grid is realised once per account
group of the client in question and the unit of a label is a (client, group,
factor) triple. A client contributing $n$ groups therefore contributes $15n$
labels, and $n$ varies by client; across the fifteen clients the corpus spans on
the order of $330$ account groups.

Table~\ref{tab:class-balance} counts at this finer level --- $4896$ labels in
total and $910$ in the test split, against the $225$ and $75$ decisions of the
grid above. The totals are not exact multiples of fifteen because assessments
falling at the framework's fourth ranking are excluded, as described above.
Accuracy, normalized mean absolute error and per-factor accuracy are all computed
over the group-level labels, so the figures of $225$ and $75$ in the main paper
describe the space of distinct decisions the task covers rather than the number of
scored predictions behind each reported metric.

\subsection{Why a Public Alternative Is Not Sufficient}
\label{sec:no-public-alternative}

What the system reads at inference is what the main paper describes: chunks of the
Financial Notes of the client's annual filing, supplied together with the inherent
risk factor to be assessed. What cannot be rebuilt from public sources is
everything around that input. The task pairs two things that are not jointly
available in public form: account-level financial detail, and a per-factor risk
level assigned by a practitioner. The filings themselves are public documents, but
the account-level trial balance that determines which account groups exist, and
therefore which assessments are in scope, is not; neither are the practitioner
assessments that serve as labels. Releasing either would disclose client
information, so the corpus is not released and the reported numbers are aggregates.

No public benchmark substitutes. Financial-NLP corpora built on public filings target
retrieval, question answering or sentiment, and none of them attaches a standardized
set of inherent risk factors, at account granularity, labelled by practitioners
applying a consistent framework. A public corpus could therefore only approximate the
surface form of the task, not the judgement being evaluated, which is precisely the
property that makes this a document-grounded transfer test rather than another
classification benchmark. That is also why we report it as a fifth domain testing
transfer, and why the paper's claims for it rest on aggregate metrics and on the
comparison against the same baseline under the same protocol, rather than on
reproducibility of the corpus itself.

\subsection{Per-Run Results and Reporting Convention}
\label{sec:per-irf-results}

The main paper states a single reporting convention --- medians over five
runs with bootstrap intervals --- and one statistic on this domain departs from it,
so we state that exception explicitly here. Accuracy is reported as the
\emph{median} over runs, as the paper describes and as used for the other domains.
Normalized mean absolute error is the exception: it is reported as the \emph{mean}.
The distinction matters only for the baseline, whose first run
attains an unusually low error: taking medians instead would give $0.369$ against
$0.305$, a gap of $0.064$ rather than the $0.045$ we report, so the mean is the more
conservative choice for our comparison.

\begin{table}[htbp]
\centering
\begin{tabular}{@{}lcccccc@{}}
\toprule
& \multicolumn{3}{c}{DGM-H} & \multicolumn{3}{c}{MetaShare} \\
\cmidrule(lr){2-4}\cmidrule(lr){5-7}
Run & Gens & Accuracy & NMAE & Gens & Accuracy & NMAE \\
\midrule
1 & 46 & 0.448 & 0.309 & 30 & 0.512 & 0.342 \\
2 & 42 & 0.445 & 0.366 & 24 & 0.544 & 0.305 \\
3 & 44 & 0.450 & 0.370 & 26 & 0.550 & 0.309 \\
4 & 45 & 0.452 & 0.371 & 27 & 0.553 & 0.302 \\
5 & 43 & 0.456 & 0.369 & 25 & 0.560 & 0.302 \\
\midrule
Reported & \textbf{44} & \textbf{0.450} & \textbf{0.357} & \textbf{26} & \textbf{0.550} & \textbf{0.312} \\
\bottomrule
\end{tabular}
\caption{Audit-risk results per run on the held-out test split. Accuracy is
aggregated as the median over runs and NMAE as the mean, as described above; lower
NMAE is better.}
\label{tab:audit-per-run}
\end{table}

\subsection{Accuracy by Risk Factor}
\label{sec:per-irf-results-factor}

Table~\ref{tab:per-factor} breaks accuracy down over the fifteen factors, in the order
of Table~\ref{tab:irf-definitions}. Each entry is the median over the five runs.

\begin{table}[htbp]
\centering
\begin{tabular}{@{}lccc@{}}
\toprule
Risk factor & DGM-H & MetaShare & $\Delta$ \\
\midrule
Account nature                 & 0.10 & 0.06 & $-0.04$ \\
Misstatement history           & 1.00 & 1.00 & $+0.00$ \\
Fraud indicators               & 0.04 & 0.57 & $+0.53$ \\
Control deficiency history     & 1.00 & 1.00 & $+0.00$ \\
Transaction volume and value   & 0.30 & 0.05 & $-0.25$ \\
Transaction complexity         & 0.01 & 0.15 & $+0.14$ \\
Transaction homogeneity        & 0.22 & 0.77 & $+0.55$ \\
Reporting complexity           & 0.42 & 0.62 & $+0.20$ \\
Loss exposure                  & 0.69 & 0.15 & $-0.54$ \\
Contingent liabilities         & 0.41 & 0.58 & $+0.17$ \\
Related-party transactions     & 0.40 & 0.60 & $+0.20$ \\
Accounting estimates           & 0.28 & 0.73 & $+0.45$ \\
Estimate process complexity    & 0.61 & 0.68 & $+0.07$ \\
Assumption subjectivity        & 0.66 & 0.61 & $-0.05$ \\
Outcome uncertainty            & 0.61 & 0.66 & $+0.05$ \\
\bottomrule
\end{tabular}
\caption{Accuracy per risk factor on the held-out test split, as the median over five
runs. $\Delta$ is MetaShare minus DGM-H.}
\label{tab:per-factor}
\end{table}

The aggregate improvement is not uniform, and the breakdown is the more informative
result. Three factors account for most of the gain, and all three require reading the
narrative disclosure rather than counting entries: transaction homogeneity
($+0.55$), fraud indicators ($+0.53$) and the presence of an accounting estimate
($+0.45$). A second group improves more modestly --- reporting complexity and
related-party transactions ($+0.20$ each), contingent liabilities ($+0.17$) and
transaction complexity ($+0.14$). Two factors are saturated for both systems, which
reflects a near-degenerate label distribution for those factors rather than a
modelling result.

Two factors regress clearly: loss exposure ($-0.54$) and transaction volume
($-0.25$). The shared memory therefore shifts \emph{where} the agent is competent
rather than lifting every factor at once, and we report the breakdown rather than the
aggregate alone because that shift is the more accurate description of what the memory
does in this domain. It also suggests a limit worth stating: the mechanism transfers
judgement that can be argued from the text of a disclosure, and does not help --- and
here actively hurts --- on factors that depend on quantities the narrative does not
contain.

\section{Polyglot: Per-Run Detail}
\label{sec:polyglot-per-run}

\subsection{Results Across the Five Runs}
\label{sec:polyglot-runs}

The main paper reports a median of $0.917$ with a $95\%$ bootstrap CI of
$[0.900, 0.933]$ over five repeated runs, reached within ten iterations.
Table~\ref{tab:polyglot-per-run} gives the runs behind that summary; with five runs
the bootstrap distribution of the median is supported on the observed values, so the
interval's endpoints fall on run values rather than between them. To
match GEA's \emph{model} configuration these runs use Claude Sonnet-4.5 as both
task- and meta-agent, which differs from the GPT~o3-mini task-agent used everywhere
else in our Polyglot results; numbers here are therefore not comparable with the
o3-mini results in Sec.~\ref{sec:polyglot-coverage} or with the main paper's
Polyglot ablation. Matching the models does not match the initialisation, as the
third limit in Sec.~\ref{sec:gea-comparison} sets out.

\begin{table}[htbp]
\centering
\begin{tabular}{@{}lcc@{}}
\toprule
Run & Best accuracy & Generations to best \\
\midrule
1 & 0.900 (54/60) & 8 \\
2 & 0.917 (55/60) & 9 \\
3 & 0.917 (55/60) & 10 \\
4 & 0.917 (55/60) & 10 \\
5 & 0.933 (56/60) & 10 \\
\midrule
Median & \textbf{0.917 (55/60)} & \textbf{10} \\
\bottomrule
\end{tabular}
\caption{Per-run Polyglot results on the small+medium set (60 tasks), with Claude
Sonnet-4.5 as both task- and meta-agent, matching the GEA configuration.
``Generations to best'' is the iteration at which the reported score first appears.}
\label{tab:polyglot-per-run}
\end{table}

Every run lands between $0.900$ (54 of 60 tasks) and $0.933$ (56 of 60), and every run
reaches its best within ten iterations.

\subsection{Comparison with GEA}
\label{sec:gea-comparison}

\begin{figure}[htbp]
  \centering
  \includegraphics[width=0.95\linewidth]{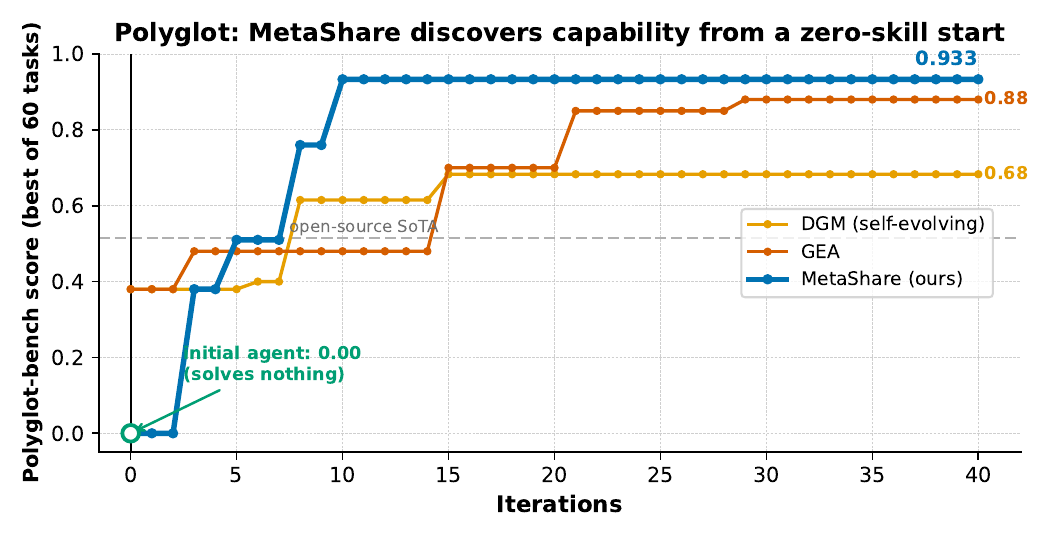}
  \caption{Best score so far against evolution iteration on the 60-task
  small+medium Polyglot subset. The MetaShare curve is the best of our five runs,
  ending at $0.933$; the median over the five is $0.917$
  (Table~\ref{tab:polyglot-per-run}). \emph{The DGM and GEA curves are digitised
  from the published figure in \citet{weng2026group}}: plateau levels and the
  iterations at which they begin were read off the plot, so they are approximate and
  indicate shape rather than exact values. Dashed line: reported open-source state of
  the art. Note that the initialisations differ---both baselines begin from an agent
  already scoring $\approx 0.38$, whereas MetaShare begins from an agent that solves
  nothing.}
  \label{fig:polyglot-comparison}
\end{figure}

On the same 60-task small+medium subset, GEA reports $0.88$ and DGM $0.683$
\citep{weng2026group}. Our median of $0.917$ and best run of $0.933$ sit
above both, and our runs reach their best within
ten iterations against roughly thirty for GEA's final plateau. Figure~\ref{fig:polyglot-comparison} places the trajectories side by
side.

Three limits should be read alongside that figure, and together they are why the main
paper describes this as competitive at a lower iteration count rather than as a win.

First, the comparison is not like-for-like on variance. GEA reports a single-run point
estimate, so there is no interval to compare against our five-run spread, and we
cannot tell whether $0.88$ is a typical or a fortunate run of that system. A gap of
$0.037$ (about two of the 60 tasks) between our median and a single unrepeated number is not evidence of a
reliable ordering.

Second, the baseline curves are digitised rather than reproduced. We did not re-run
GEA or DGM; we read their plateaus off the published figure. That is enough to compare
the shape of the trajectories and the iteration at which each plateaus, but the
individual values carry reading error, and small differences between the curves should
not be interpreted.

Third, the runs do not share an initialisation. The main paper's ``matching that
configuration'' refers to the models each system runs; the starting agent is not
matched. GEA and DGM are seeded with an agent
that already solves roughly $38\%$ of the subset, while our agent starts at zero and
must discover every capability the curve shows. This difference runs against us in the
final-score comparison, since our system covers the same ground from further back, but
it also means the two curves are not measuring the same quantity over the same
interval: the early part of our trajectory is capability acquisition, whereas for the
baselines it is refinement of an already competent agent.

\subsection{Task Coverage Across Generations}
\label{sec:polyglot-coverage}

Peak accuracy is a property of a single checkpoint, and it can hide what the search
actually did. Two runs can solve the same \emph{number} of tasks in every generation
and still differ in whether those are the same tasks each time or different ones;
only the second is exploring. We therefore also report \emph{cumulative unique tasks
solved}: we take each generation, record which coding tasks its agent solves, and
count the distinct tasks solved at any point in the run, over the 60-task
small+medium subset of the 225-task Polyglot benchmark. Because it accumulates over
the whole run and is insensitive to how often a task is re-solved, this quantity
measures the variety the population reaches rather than the quality of one endpoint,
and it upper-bounds the task set of any individual checkpoint. It is a different
measurement from the best-so-far accuracy in the main paper's Polyglot ablation,
which is taken on the full 225-task benchmark at the best-performing generation; the
two are not commensurable and should not be read against each other.

Both arms receive the same 100-generation budget and the same staged evaluation
protocol, and in both the task-agent runs on GPT~o3-mini for the whole run, with
Claude Sonnet-4.5 as meta-agent. This is also what the caption of Figure~\ref{fig:main-comparison} of the
main paper means by holding the task-agent on o3-mini for Polyglot: the runs
reported there are ones in which the task-agent remained on o3-mini for the whole
run, not ones in which switching was disabled. Model switching
is in the action space of both arms, as elsewhere on this domain, and was neither
forced nor blocked in either of them; the five MetaShare runs reported here are ones
in which the meta-agent did not exercise it, so the coverage comparison isolates
the search rather than a difference in task-agent backbone. The runs in which
routing does emerge are traced separately in Sec.~\ref{sec:emergent-routing} and
are not counted here, so this table understates the full system rather than
flattering it.

\begin{table}[htbp]
\centering
\begin{tabular}{@{}lccc@{}}
\toprule
Run & DGM-H & MetaShare & $\Delta$ \\
\midrule
1 & 21 & 27 & $+6$ \\
2 & 22 & 26 & $+4$ \\
3 & 21 & 27 & $+6$ \\
4 & 22 & 27 & $+5$ \\
5 & 21 & 24 & $+3$ \\
\midrule
Median & \textbf{21} & \textbf{27} & $+6$ \\
Mean   & 21.4 & 26.2 & $+4.8$ \\
\bottomrule
\end{tabular}
\caption{Cumulative unique Polyglot tasks solved per run, out of the 60-task
small+medium subset, over a matched 100-generation budget.}
\label{tab:polyglot-coverage}
\end{table}

MetaShare covers more tasks in every run, and the two sets of run outcomes do not
overlap: the baseline spans 21--22 tasks while MetaShare spans 24--27. The median
rises from 21 to 27 of the 60 tasks, a relative increase of $1.29\times$ on medians
and $1.22\times$ on means. The separation is small in absolute terms but consistent,
and it is not bought with extra generations: both arms run the full 100 generations,
and MetaShare's best checkpoint arrives well before that budget is exhausted.

We read this as evidence for the mechanism rather than for the headline score. Under
failure-aware sharing, a branch that has recorded a dead end does not need to
rediscover it, and the signed records make that dead end legible to genealogically
distant branches; the search therefore spends its budget on tasks it has not yet
attacked instead of re-attempting known-bad mechanisms. Coverage is the measurement
where that effect should appear most directly, since it rewards reaching new tasks
rather than refining an existing solution. Two caveats bound the claim. The counts are
per-run and not pooled, so they are not a union over the five runs and should not be
added. And coverage credits a task solved at any generation, so it does not certify
that the finally selected checkpoint solves all of the tasks counted.

\subsection{Emergent Dynamic Model Selection}
\label{sec:emergent-routing}

The main paper reports that on Polyglot the auditor unlocks a capability rather than
an incremental gain: the meta-agent discovers that it can change the task-agent's
model, which DGM-H never proposes despite having the same action available. This
section traces that episode through one Polyglot run that begins on GPT~o3-mini
and leaves it at generation~20, from the search collapse that triggered the audit
to the modification that followed. The directive
and pool digest reproduced below are one sample, drawn from a single audit in that
run and shown for reference; both the categories proposed and the records retrieved
differ across runs and across audits within a run.

\paragraph{The search collapses before this audit.} The audit traced here is the one
that fires at generation~20, over a pool of nineteen records; under the cadence $C$
of Table~\ref{tab:hyperparameters} it need not be the first audit of the run, and it
is singled out because its directive is the one the meta-agents then acted on.
Table~\ref{tab:audit-episode}
summarises the mechanism types the classifier assigned to the pool records of this
run. Over the first nineteen generations the population produced exactly two of the
six mechanism types, and the concentration is stronger than those counts suggest:
generations~10 through~19 are ten consecutive records that all implement
self-consistency decoding, contributed by seven distinct lineages. This is the
failure mode the auditor exists to detect, and it is a population-level failure
rather than one branch getting stuck --- independent lineages converged on the same
recipe. Every one of the nineteen records carries a performance delta of $0.000$: the
run had not yet solved anything.

\begin{table}[htbp]
\centering
\begin{tabular}{@{}lcc@{}}
\toprule
Mechanism type & Gen.\ 1--19 (pre-audit) & Gen.\ 20--24 (post-audit) \\
\midrule
\texttt{evaluation\_logic}    & 12 & 1 \\
\texttt{prompt\_engineering}  & 7  & 0 \\
\texttt{tool\_creation}       & 0  & 1 \\
\texttt{structural\_refactor} & 0  & 1 \\
\texttt{memory\_system}       & 0  & 0 \\
\texttt{self\_modification}   & 0  & 0 \\
\midrule
Distinct types & 2 & 3 \\
\bottomrule
\end{tabular}
\caption{Mechanism types in the pool of one Polyglot run that begins on GPT
o3-mini, before and after the generation-20 audit fired. Types are assigned by the post-hoc classifier of
Sec.~\ref{sec:prompt-classifier}; the six available types are listed in
Table~\ref{tab:strategy-types}.}
\label{tab:audit-episode}
\end{table}

\paragraph{The directive.} The auditor sees the nineteen diffs, with evaluation
content removed by the scrubbing of Sec.~\ref{sec:leakcheck}, and returns the text in
Prompt~\ref{prompt:audit-output}. It names the redundancy it found and proposes
eight absent categories. Nothing in the returned text refers to a task, a test case,
or a score; the reasoning is entirely about the shape of the diffs it was shown.

\begin{lstlisting}[style=promptAudit,caption={\paudit{} Directive returned by the
auditor for this run, verbatim. Dynamic Model Selection is the first of eight
proposed categories. Shown as one sample directive; the categories proposed vary
between audits.},label={prompt:audit-output}]
Looking at the 19 diffs, I see heavy emphasis on prompt engineering and
self-consistency voting mechanisms, but several important categories are
missing or under-represented:

1. **Dynamic Model Selection**
   - Automatically switches between different models or model configurations
     based on task characteristics or intermediate results. Introduces model
     routing logic and performance tracking across model variants.

2. **Hierarchical Problem Decomposition**
   - Breaks complex tasks into subtasks with recursive solving and result
     composition. Structurally different by introducing task splitting
     functions and dependency management between subtask solutions.

3. **Memory-Based Learning from Failures**
   - Maintains persistent storage of past failures and successful strategies
     to inform future attempts. Adds failure pattern recognition and strategy
     adaptation based on historical performance data.

4. **Multi-Agent Collaboration**
   - Uses multiple specialized agent instances with different roles (e.g.,
     critic, solver, validator) that interact to reach solutions. Introduces
     agent coordination protocols and role-based reasoning chains.

5. **Adaptive Timeout and Resource Management**
   - Dynamically adjusts computation time, tool usage limits, or reasoning
     depth based on task complexity and available resources. Adds resource
     monitoring and adaptive constraint adjustment mechanisms.

6. **Iterative Solution Refinement**
   - Generates initial solutions then systematically improves them through
     multiple refinement cycles. Introduces solution evolution loops and
     incremental improvement tracking, distinct from simple retry logic.

7. **Confidence-Weighted Ensemble Methods**
   - Combines multiple solution approaches using learned confidence weights
     rather than simple voting. Adds confidence calibration and weighted
     aggregation beyond basic consensus mechanisms.

8. **Dynamic Prompt Optimization**
   - Automatically modifies prompts based on intermediate results or detected
     failure modes. Introduces prompt mutation and selection mechanisms that
     evolve during execution.
\end{lstlisting}

\paragraph{How the directive reaches the meta-agent.}
Prompt~\ref{prompt:assembled-context} shows the assembled context, with the
directive at the top as a mandatory instruction and the pool digest below it as
reference material. Three features of the assembly are visible here. The directive is
cached: the text says ``the 19 diffs'' while the digest beneath it already contains
the generation-20 record, because the audit output is carried forward until the next
audit fires rather than regenerated every generation. The success and failure blocks
are separate, and only one positive record appears because only one existed --- the
retrieval budget $K_s$ is an upper bound, not a quota. The failure block's heading is
the implementation's own wording, and it names both cases: the operative criterion is
the non-positive test $\delta_i \leq 0$, so records with $\delta_i = 0.000$ appear
there as strategies that did not work rather than as ones that reduced
performance. And the
unexplored-directions
footer is computed from the classifier's type inventory, so it lists the two types
still absent from the pool at that point.

\begin{lstlisting}[style=promptAudit,caption={\paudit{} Assembled meta-agent context
in a generation after the routing change was recorded. Strategy bodies are elided
where they repeat; the directive block is the cached output of
Prompt~\ref{prompt:audit-output}. Shown as one sample
context.},label={prompt:assembled-context}]
--- MUST TRY: mechanism categories absent from all prior strategies (eval-blind audit) ---
You MUST attempt one of the categories below. Do not produce another variation
of the recipes shown later -- that path has plateaued.

[the eight-category directive above is reproduced here verbatim]

--- PRIOR STRATEGIES (reference only -- do NOT clone the dominant recipe) ---

[Strategy 1] +0.500 on polyglot (gen_20, branch 13):
  This change adds a dynamic model routing system that analyzes task
  characteristics to automatically select the most appropriate language model
  for each task, then tracks performance metrics (success rate and execution
  time) to improve future model selection decisions. [...]

--- STRATEGIES THAT REDUCED PERFORMANCE or DID NOT WORK (avoid these approaches)---

[Strategy 2] +0.000 on polyglot (gen_1, branch 0):
  This change enhances the meta-agent's self-improvement by providing it with
  concrete guidance on what code areas to modify (task agent, tools, LLM
  integration, etc.) and incorporating recent evaluation results as context
  for deciding what improvements to make. [...]

[Strategy 3] +0.000 on polyglot (gen_2, branch 0):
  This change enhances the agent's prompt by adding explicit identity framing
  ("capable problem-solving agent"), encouraging step-by-step reasoning [...]

[Strategies 4-6: three further prompt-scaffolding variants, gen_3, gen_5, gen_6]

--- UNEXPLORED STRATEGY DIRECTIONS: memory_system, self_modification ---
\end{lstlisting}

\paragraph{What the meta-agents proposed.} In the generation following the audit, all
three parallel meta-agents abandoned the self-consistency recipe and each targeted a
category from the directive. Two proposed Dynamic Model Selection, creating
\texttt{agent/model\_router.py} and \texttt{agent/model\_selector.py} respectively;
the third proposed Hierarchical Problem Decomposition, creating
\texttt{agent/task\_decomposer.py}. All three cleared every axis of the admissibility
gate of Sec.~\ref{sec:gate} with a composite score of $1.0$, so the three-way tie was
resolved by the generation-seeded tie-break draw of Sec.~\ref{sec:gate}, which in
this generation selected the first candidate. The resulting
record is the only positive entry in the run's pool, moving the score from $0.000$ to
$0.500$. That score is the training-time score of this single run on the 60-task
small+medium subset, which is what the evolution loop scores against; it is neither an
aggregate over the five runs nor a figure on the 225-task benchmark, so it should not
be read against the Polyglot accuracies of the main paper. It is also measured after
the routing change, with GPT~4o-mini serving the tasks, so it is not comparable with
the coverage counts of Sec.~\ref{sec:polyglot-coverage} either, which come from runs
whose task-agent stays on o3-mini throughout.

\paragraph{What the modification actually does.} The selected patch adds a router that
classifies each task and selects the model the task-agent runs it on, recording
per-model success rate and execution time to inform later routing. In this run the
router sent Polyglot tasks to GPT~4o-mini, so in practice the task-agent moved off
o3-mini to GPT~4o-mini and stayed there --- exactly the model change the main paper
reports. The original model is restored after each task, so o3-mini survives as the
fallback default when the router declines to route. This upgrade is the task-agent
model change that the main paper identifies as an uncontrolled difference between the
arms on this domain, and the control reported there applies unchanged: the runs that
stay on o3-mini throughout still beat the DGM-H baseline in fewer iterations.

\paragraph{Contrast and caveats.} DGM-H has the same action available and does not
take it in any run, which is the observation the ablation rests on. One limit applies:
the episode above is a single run traced in detail, offered as a mechanism account
rather than as evidence of frequency; the reproducibility claim rests on the ablation
arms, not on this trace.

\end{document}